\documentclass[pra,aps, twocolumn, superscriptaddress, longbibliography, notitlepage, floatfix]{revtex4-2}
\usepackage[utf8]{inputenc} 
\usepackage{graphicx}
\usepackage{bm}
\usepackage{dsfont}
\usepackage{amsmath,amsfonts, amssymb, chngcntr, lipsum}
\usepackage[english]{babel}
\usepackage{amsthm}
\usepackage{marvosym}
\usepackage[normalem]{ulem}
\usepackage{xcolor}
\usepackage{color,soul}
\usepackage{tikz}
\usepackage[colorlinks=true, citecolor=blue, urlcolor=blue, allcolors=blue]{hyperref}

\definecolor{newtext}{RGB}{1.0, 0.0, 0.0}

\begin{document}

\title{Heat operator approach to quantum stochastic thermodynamics in the strong-coupling regime}

\author{Sheikh Parvez Mandal}
\email{sheikhparvez.mandal@um.es}
\affiliation{Departamento de Física - CIOyN, Universidad de Murcia, Murcia E-30071, Spain}
\author{Mahasweta Pandit}
\affiliation{Departamento de Física - CIOyN, Universidad de Murcia, Murcia E-30071, Spain}
\author{Khalak Mahadeviya}
\affiliation{School of Physics, Trinity College Dublin, College Green, Dublin 2, D02 K8N4, Ireland}
\author{Mark T. Mitchison}
\email{mark.mitchison@kcl.ac.uk}
\affiliation{School of Physics, Trinity College Dublin, College Green, Dublin 2, D02 K8N4, Ireland}
\affiliation{Department of Physics, King’s College London, Strand, London, WC2R 2LS, United Kingdom}
\author{Javier Prior}
\email{javier.prior@um.es}
\affiliation{Departamento de Física - CIOyN, Universidad de Murcia, Murcia E-30071, Spain}
  
\begin{abstract}
\textcolor{newtext}{Heat exchanged between an open quantum system and its environment exhibits fluctuations that carry crucial signatures of the underlying dynamics. Within the well-established two-point measurement scheme, we identify a `heat operator,' whose moments with respect to the vacuum state of a thermofield-doubled Hilbert space correspond to the stochastic moments of the heat exchanged with a bath. This recasts heat statistics as a unitary time evolution problem, which we solve by combining chain-mapped reservoirs with tensor network propagation. In a multi-bath setup all total and bath-resolved heat moments then follow from a single pure state evolution. We employ this approach to compute transient and steady state heat fluctuations in Ohmic spin-boson models in and out of equilibrium, accessing the challenging low temperature and long memory time regimes of the environment. In the nonequilibrium case, we show a crossover in the Fano factor from super-Poissonian to nearly Poissonian statistics under strong coupling asymmetry, corresponding to thermal rectification behavior. The method applies to noninteracting (bosonic or fermionic) nonequilibrium environments with arbitrary spectral densities, offering a powerful, non-perturbative framework for understanding heat transfer in open quantum systems.}
\end{abstract}

\maketitle

In open quantum systems, heat exchanged with the environment is not only a fundamental source of decoherence~\cite{Popovic2023} and dissipation~\cite{Landi2021,Strasberg2021} but can also be exploited for technological benefit, e.g.~to drive useful autonomous machines~\cite{aamir_thermally_2025,AntonioMarinGuzman2024} or characterize unknown noise sources~\cite{Spiecker2023}. 
Moreover, quantum coherence can induce non-classical heat fluctuations~\cite{levy_quasiprobability_2020, Kerremans2022}, whose important consequences range from dramatically increased dissipation during the erasure of quantum information~\cite{Miller2020} to 
reduced power fluctuations in quantum thermal machines~\cite{ptaszynski_coherence-enhanced_2018,agarwalla_assessing_2018,brandner_thermodynamic_2018,guarnieri_thermodynamics_2019,kalaee_violating_2021,rignon-bret_thermodynamics_2021}. However, in general, open quantum systems are significantly affected by their surroundings due to their small size, leading to complex phenomena including non-Markovian dynamics~\cite{rivas_quantum_2014,breuer2016}, level broadening~\cite{josefsson_quantum-dot_2018,Josefsson2019}, and modified equilibrium states~\cite{talkner_colloquium_2020,Trushechkin2022}, which are absent from conventional thermodynamic frameworks \cite{talkner_colloquium_2020}. Predicting fluctuating heat exchange in open quantum systems with strong system-environment coupling is thus an important yet challenging problem, driving the recent development of advanced computational methods~\cite{Wang2014, Carrega2016,Segal2016,Song2017,kilgour_path-integral_2019,Yadalam2022}.

Tensor network algorithms provide powerful tools to simulate open quantum system dynamics in the strong-coupling regime. Various strategies exist, including mapping the environment onto a chain that is amenable to exact simulation~\cite{prior_efficient_2010,prior2013quantum, Tamascelli2019,Nuesseler2020}, representing the environment's influence by a process tensor~\cite{strathearn_efficient_2018,jorgensen_exploiting_2019,Thoeniss2023,cygorek_simulation_2022}, or approximating the environment by a collection of damped pseudomodes~\cite{Somoza2019, Wojtowicz2020, Lotem2020, Brenes2020, Purkayastha2021, Cirio2024}. 
However, none of the aforementioned methods can be straightforwardly used to predict heat fluctuations, because heat is not a state function (i.e., an observable). Rather, heat is a process-dependent quantity that depends on the outcomes of (at least) two measurements at different times~\cite{esposito_nonequilibrium_2009,talkner_colloquium_2020, aurell2018characteristic}. Recent works have sought to solve this problem in the strong-coupling regime, either by targeting the characteristic function while sacrificing the trace-preserving character of the evolution~\cite{popovic_quantum_2021,Brenes2023,Shubrook2025,Valli2024}, or by exploiting a Markovian embedding~\cite{woods_mappings_2014,Tamascelli2018} to perform stochastic unraveling of the dynamics~\cite{Menczel2024,Bettmann2024}, at the cost of sampling a large number of trajectories. These difficulties are compounded for environments with complex spectral features or low temperatures, since more pseudomodes or higher bond dimensions are requinewtext to capture the associated non-Markovian effects.

\begin{figure}
    \centering
        \includegraphics[ trim= 5.2cm 17.8cm 5.5cm 4cm, clip, width=.85\linewidth]{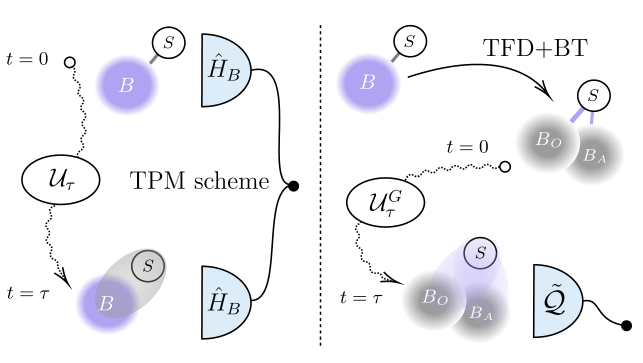}
    \caption{In the two-point measurement (TPM) scheme for calculating heat moments, the bath Hamiltonian $\hat H_B$ is measured at times $t=0$ and $\tau$. The joint system-bath ($S$-$B$) state evolves under $\mathcal{U}_t$. In the proposed method (right), thermofield doubling (TFD) and Bogoliubov transformation (BT) take a Gaussian state on $B$ to a vacuum state on $B_O$-$B_A$. Measuring the `heat operator' $\tilde{\mathcal{Q}}$ at $t=\tau$ on the unitarily evolved joint $B_A$-$S$-$B_O$ state then yields the heat moments of $B$. 
    }
    \label{fig:front_page}
\end{figure}

\textcolor{newtext}{In this work, we calculate quantum heat statistics by introducing a `heat operator', $\tilde{\mathcal{Q}}$, which acts on an extended Hilbert space (Fig.~\ref{fig:front_page}) that furnishes a purification of the environment's initial thermal state. Remarkably, our main results (Eqs.~\eqref{eq:simpl_chi_q} and \eqref{eq:main_result}) recast two-point measurement of heat statistics as single-time expectation values, albeit with a modified (yet still unitary) time evolution starting from a vacuum state of the extended environment. We use the proposed framework to capture rectification of heat current and its fluctuations on an equal footing, in both transient and non-equilibrium steady states of a two-level system coupled to low temperature and highly non-Markovian bosonic environments. These results are directly applicable to describe contemporary quantum physics experiments, for example, in superconducting circuits, trapped ions, solid state spins, and other platforms featuring strong coupling to structured environments.}

\textit{Setup.---} We consider an open quantum system $S$ with Hamiltonian $\hat{H}_S(t)$, whose time dependence takes into account any non-dissipative control or driving on $S$ by an external agent. The environment comprises some number $N$ of baths, $B_j$, where $j=1,\ldots,N$. These baths, which can either be bosonic or fermionic, are described by free Hamiltonians $\hat{H}_{B,j}=\sum \nolimits _{\nu} \omega_{j\nu} \hat{a}^{\dagger}_{j\nu} \hat{a}_{j\nu}$. We will take $\hbar=1$. The mode operators $\hat{a}_{j\nu}$ satisfy the standard canonical bosonic or fermionic algebra. We assume that $S$ couples independently to each $B_j$ through a linear interaction Hamiltonian
\begin{equation}
    \label{eq:interaction_ham}
    \hat{H}_{I,j}=\sum_{\nu} g_{j\nu} \hat{L}^{\dagger}_j \hat{a}_{j\nu} + \text{h.c.}, 
\end{equation}
with coupling constants $g_{j\nu}$ and arbitrary system operators $\hat{L}_j$, leading to the total system-bath Hamiltonian
\begin{equation}
    \hat{H}(t) = \hat{H}_{S}(t)+  \sum_{j}  \hat{H}_{I,j} + \hat{H}_{B,j}.  \label{eq:original_H}
\end{equation}
In this setting, the effect of each bath on the system's evolution is characterized by the spectral density, $J_j(\omega) = 2\pi \sum\nolimits_{\nu} |g_{j\nu}| ^2 \ \delta(\omega - \omega_{j\nu})$, which becomes a continuous function in the limit of an infinite bath and a smooth density of states \cite{weiss_quantum_2012, breuer_theory_2007}. We assume that the baths are initially in thermal equilibrium and uncorrelated with the system, leading to the initial global density matrix $\hat{\rho}(0)  = \hat{\rho}_S(0) \otimes \hat{\pi}_{\beta_1} \ldots \otimes\hat{\pi}_{\beta_N}$, where $\hat{\pi}_{\beta_j}$ is a thermal state of bath $j$ at inverse temperature $\beta_j$. The state then evolves as $\hat{\rho}(t) = \mathcal{U}_t[\hat{\rho}(0)] \equiv \hat{U}(t) \hat{\rho}(0) \hat{U}^\dagger(t)$, where $\hat{U}(t)$ is the unitary operator generated by Eq.~\eqref{eq:original_H}. Heat, however, is not a state function, being defined by the two-point measurement (TPM) described below.

\textit{Quantum heat statistics.---}Following standard nomenclature~\cite{Landi2021,Strasberg2021}, we identify the energy change of each bath as heat; however, we note that other approaches to heat at strong coupling~\cite{talkner_colloquium_2020,Colla2022} have been proposed in the recent literature~\footnote{Our approach aligns with the framework of Ref.~\cite{talkner_colloquium_2020} so long as the system-environment interaction is assumed to be switched on and off at the beginning and end of the evolution time $t$, which is fully consistent with our assumption of a product initial condition.}. For the sake of simplicity, we consider a single bath, i.e.~$N=1$, and omit the redundant index $j$ in the rest of the paper, but our approach generalizes straightforwardly to multiple baths. The heat $Q(t)$ is a stochastic variable defined by a projective two-point measurement of the bath Hamiltonian $\hat{H}_{B}$ at the initial time $0$ and final time $t$~\cite{esposito_nonequilibrium_2009, talkner_colloquium_2020}. We decompose $\hat{H}_{B}$ as $\sum_n E_{n}\hat{\Pi}_{n}$, where $\{\hat\Pi_{n}\}$ is the unique family of projection operators corresponding to the eigenvalues $\{E_{n}\}$ of $\hat{H}_{B}$. Born's rule gives us the joint probability of fluctuation from $E_{m}$ to $E_{n}$ to be $P[E_{n}, E_{m}]  = \text{Tr} \{ \hat{\Pi}_{n} \mathcal{U}_t[\hat{\Pi}_{m} \hat{\rho}(0)\hat{\Pi}_{m}]\}$. The probability distribution of heat fluctuations is then given by $P(Q,t)  = \sum_{m,n} P[E_{n}, E_{m}]\delta (Q + E_{m} - E_{n})$. Using this distribution, we determine the $n^{\rm th}$ moment of $Q(t)$ as
\begin{equation}
    \langle Q^n(t)\rangle = \int d Q P(Q,t) Q^n  = (-i)^n \partial^n_\lambda \chi(\lambda,t) \big|_{\lambda=0}, \label{eq:mu_moments}
\end{equation}
where $\chi(\lambda, t)=\int dQ\ P(Q, t)\, e^{i\lambda Q}$ is called the characteristic function of $P(Q,t)$ and $\lambda$ is the counting field. 
The characteristic function uniquely determines the probability distribution $P(Q, t)$, and can be expressed as follows (see Appendix \ref{app:chiu_q})
\begin{equation}
    \chi(\lambda,t) = \text{Tr}\left\{e^{i\lambda\hat H_{B}} \,\mathcal{U}_t[e^{-i\lambda\hat H_{B}} \hat{\rho}(0) ]\right\}\label{eq:chiu_q}.
\end{equation}
Thus, we recover the first moment as the mean energy change of the bath, $\langle Q(t))\rangle = \text{Tr}\{\hat H_{B}(\hat{\rho}(t)-\hat{\rho}(0))\}$. The second moment $\langle Q^2(t) \rangle$ is related to the variance of heat as $\langle \langle Q^{2}(t)\rangle \rangle = \langle Q^2(t) \rangle - \langle Q(t) \rangle^2$, and so on. Unlike the first moment, however, higher moments do not have a simple expression in terms of quantum expectation values. Therefore, an efficient procedure to evaluate $\chi(\lambda,t)$ or its derivatives is crucial, and we describe our method below.

\textcolor{newtext}{\textit{Heat-operator approach}}.---The form of Eq.~\eqref{eq:chiu_q} \textcolor{newtext}{obtained using the TPM scheme} poses two apparent obstacles to a direct numerical evaluation of $\chi(\lambda,t)$. The first stems from initialization of the bath's thermal state $\hat \pi_\beta$, which can be achieved by imaginary time evolution starting from a maximally mixed state. This procedure is costly for certain representations of the bath modes, including the 1D chain geometry pertinent for the chain mapping of baths using orthogonal polynomials~\cite{prior_efficient_2010}. In the language of tensor networks, imaginary-time evolution generates larger bond dimension (which is a measure of correlations) of the matrix-product operator (MPO)~\cite{zwolak_mixed-state_2004, verstraete_matrix_2004, Javier09, mcculloch_infinite_2008} representing the bath state, especially while targeting low, nonzero temperatures. Second, the propagator in Eq.~\eqref{eq:chiu_q} is dressed by the counting field via the operators $e^{\pm i \lambda H_B}$, leading to a non-unitary (trace-decreasing) overall time evolution, which can be unstable for standard tensor-network methods, especially for higher heat moments.

We sidestep both these issues using the thermofield approach \cite{takahashi_thermo_1996,das_supersymmetry_1989} that introduces a decoupled (otherwise identical) auxiliary environment (denoted by $A$) to the original system-bath Hamiltonian (denoted as $S$ and $O$ respectively) \cite{karrasch_finite-temperature_2012, de_vega_thermofield-based_2015}; see Fig.~\ref{fig:front_page} for an illustration. Bath $A$ consists of modes $\hat{b}_{\nu}$ of frequency $-\omega_\nu$ that correspond to every mode $\hat{a}_{\nu}$ of frequency $\omega_\nu$ in bath $O$. This effectively doubles the Hilbert space of the bath, akin to a purification scheme. \textcolor{newtext}{The auxiliary bath can be interpreted as a quantum register that coherently stores the outcome of the first measurement, in accordance with the principle of deferred measurement~\cite{nielsenQuantumComputationQuantum2012}.} The total Hamiltonian in \eqref{eq:original_H} is then transformed to, 
\begin{equation}
    \Tilde{H}(t) =  \hat{H}(t) - \hat H_{B,A},\quad \hat H_{B,A} =  \sum \nolimits _{\nu} \omega_{\nu} \hat{b}^{\dagger}_{\nu} \hat{b}_{\nu}.  \label{eq:HTF}
\end{equation}
The initial state of bath $O$ which is a Gaussian state, $\hat{\pi}_{\beta}$,
 then gets mapped to a pure state in the extended Hilbert space of $OA$ (original + auxiliary), referred to as the `thermofield double' or `thermal vacuum' state: $\Tilde{\tau}_\beta=|\psi(\beta)\rangle_{\text{\tiny $OA$}}\langle \psi(\beta)|_{\text{\tiny $OA$}}$. The thermofield transformation ensures that 
\begin{equation}
    \text{Tr}_{A}\{\Tilde{\tau}_\beta\}= \hat{\pi}_{\beta}=\frac{e^{-\beta (\hat{H}_{B}-\mu \hat N)}}{\text{Tr}\{e^{-\beta (\hat{H}_{B}-\mu \hat N)}\}} . \label{eq:trace_tau}
\end{equation}
Here, for generality, we have included a chemical potential $\mu$ coupled to the particle number $\hat{N}$ of the bath. Purification, which is a linear operation, preserves the Gaussianity of the state $\Tilde{\tau}_\beta$. Now, there always exists a Bogoliubov transformation \cite{bogoljubov_new_1958, balian_nonunitary_1969, nam_diagonalization_2016}, $\mathcal{G}_{\text{\tiny$\beta$}}$, which squeezes $\Tilde{\tau}_\beta$ to the vacuum state of $OA$: 
\begin{equation}
\Tilde{\Phi} = \mathcal{G}_{\text{\tiny$\beta$}}[\Tilde{\tau}_{\text{\tiny$\beta$}}] = \bigotimes_\nu\, \phi_{\nu,O}\otimes  \phi_{\nu,A} =  |\Tilde{\Phi}\rangle \langle \Tilde{\Phi}|,  \label{eq:vacuum_OA}  
\end{equation}
where we denote the zero occupation state of mode $\nu$ as $\phi_\nu=|0_\nu\rangle \langle 0_\nu|$ \textcolor{newtext}{and $|\Tilde{\Phi}\rangle= \bigotimes_\nu |0_{\nu,O}\rangle|0_{\nu,A}\rangle$ is the initial vacuum state of baths $OA$}.\ The transformation $\mathcal{G}_\beta$ is canonical, i.e.~it preserves the bosonic or fermionic (anti-)commutation relations, and it can be shown \cite{martino_quantum_1996, celeghini_thermo_1998} that $\mathcal{G}_\beta[\bullet] = e^{-i\hat{G}_\beta}\bullet e^{i\hat{G}_\beta}$ is generated by the Hermitian operator $\hat G_{\text{\tiny$\beta$}} = - i \sum \nolimits _{\nu} \theta_{\text{\tiny$\beta$},\nu} (\hat{a}_{\nu}^{\dagger}\hat{b}_{\nu}^{\dagger}-\hat{b}_{\nu}\hat{a}_{\nu})$, \textcolor{newtext}{with $\theta_{\text{\tiny$\beta$},\nu}$ depending on the thermal occupation of the $\nu^{\rm th}$ mode, $ n_{\beta,\nu} = [e^{\beta(\omega_\nu-\mu)}+(-1)^\eta]^{-1}$ as shown below}. This transformation mixes the modes of the baths $O$ and $A$ as follows:
\begin{align}
    &\mathcal{G}_{\text{\tiny$\beta$}}[\hat a_{\nu} ]  = u_{\text{\tiny$\beta$},\nu} \hat a_{\nu} + v_{\text{\tiny$\beta$},\nu} \hat b_{\nu}^{\dagger},  \nonumber  \\ &\mathcal{G}_\beta[\hat b_{\nu} ]  = u_{\text{\tiny$\beta$},\nu} \hat b_{\nu} + (-1)^ {\eta +1} v_{\text{\tiny$\beta$},\nu} \hat a_{\nu}^{\dagger},    
\end{align}
with $v_{\text{\tiny$\beta$},\nu} =   \text{sin}((-i)^ \eta \theta_{\text{\tiny$\beta$}, \nu}) = \sqrt{n_{\text{\tiny$\beta$},\nu}}$ and $u_{\text{\tiny$\beta$},\nu} ^2 + (-1)^\eta v_{\text{\tiny$\beta$},\nu}^2 =1$, $u_{\text{\tiny$\beta$},\nu} \geq 0$. Here, $\eta = 0$ or $\eta =1$ for fermions or bosons, respectively.


\textit{Main result.---}We now introduce the `heat operator' defined on $OA$,
\begin{equation}
    \Tilde{\mathcal{Q}}:= \hat H_{B,O}-\hat H_{B,A}. \label{eq:heat_operator}
\end{equation} 
For any Gaussian initial bath state, 
\begin{equation}
    \chi (\lambda,t)  = \text{Tr}\left\{e^{i\lambda \Tilde{\mathcal{Q}}}\,\mathcal{U}^{\text{\tiny $G$}}_t[\hat \rho _{\text{\tiny $S$}}(0) \otimes \Tilde{\Phi}]\right\}. \label{eq:simpl_chi_q}
\end{equation}
Here, $\mathcal{U}^{\text{\tiny $G$}}_t[\bullet] = \tilde{U}_{\text{\tiny $G$}}(t) \bullet \tilde{U}_{\text{\tiny $G$}}^\dagger(t)$ and $\tilde{U}_{\text{\tiny $G$}}(t)$ is the unitary time evolution operator generated by $\Tilde{H}_{\text{\tiny $G$}}(t)=\mathcal{G}_{\text{\tiny$\beta$}}[\Tilde{H}(t)]$, which is given explicitly by
\begin{equation}\label{transformed_Hamiltonian}
\Tilde{H}_{\text{\tiny $G$}}(t)= \Tilde{H}(t) + \sum_\nu \left(u_{\text{\tiny$\beta$},\nu} g_{\nu}\hat{L}^{\dagger} \hat{a}_{\nu} + v_{\text{\tiny$\beta$},\nu} g_{\nu} \hat{b}_{\nu}\hat{L}  + \text{h.c.}\right). 
\end{equation}
\textcolor{newtext}{\textit{Proof sketch.---}Starting from Eq.~\eqref{eq:chiu_q}, we purify the thermal bath state $\hat{\pi}_\beta$
by performing the thermofield transformation which introduces bath Hamiltonian $\hat H_{B,A}$ (Eq.~\eqref{eq:HTF}), the pure state $\tilde{\tau}_\beta$ (Eq.~\eqref{eq:trace_tau}), and the heat operator $\Tilde{\mathcal{Q}}$ (Eq.~\eqref{eq:heat_operator}). Using the facts that $\Tilde{\mathcal{Q}}$ commutes with $\hat{G}_{\text{\tiny$\beta$}}$
and that it also annihilates $\hat{\rho}_{\text{\tiny $S$}}(0)\otimes \tilde{\Phi}$, we show that the $\chi (\lambda,t)$ reduces to the single-time expectation value in Eq.~\eqref{eq:simpl_chi_q}, with evolution generated by $\tilde{H}_G(t)$. The details of the proof are presented in Appendix~\ref{app:state_proof}.} 

It follows from Eq.~\eqref{eq:simpl_chi_q} that 
\begin{equation}
       \langle Q ^n(t)\rangle = \langle\Tilde{\mathcal{Q}}^{n}(t)\rangle \equiv \text{Tr}\left\{\Tilde{\mathcal{Q}}^n \, \mathcal{U}^{\text{\tiny $G$}}_t[\hat \rho _{\text{\tiny $S$}}(0) \otimes \Tilde{\Phi}]\right\},\label{eq:main_result}
\end{equation}
which states that each moment of stochastic heat equals the corresponding moment of the heat operator $\tilde{\mathcal{Q}}(t)$, where the time evolution is generated by the transformed Hamiltonian $\tilde{H}_G(t)$ and the expectation value is taken with respect to an initial vacuum state $\tilde{\Phi}$ of the extended bath $OA$. \textcolor{newtext}{Thus, the requirement of non-unitary preparation and evolution steps with a single bath $O$ in the conventional scheme (Eq.~\eqref{eq:chiu_q}) is replaced by pure unitary evolution of an extended zero temperature bath $OA$. This is particularly advantageous for low temperatures, multiple bath settings, and mitigates error accumulation in the higher-order heat cumulants by avoiding non-unitary evolution.}

\begin{figure}[t]
    \centering
    \includegraphics[trim= 5.2cm 20.cm 7cm 4.35cm, clip, width=.94\linewidth]{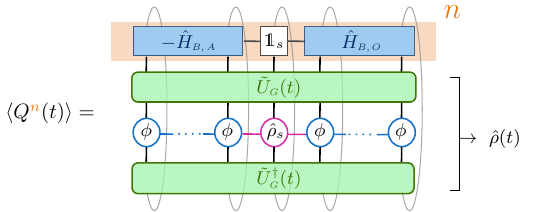}
    \caption{Diagram illustrating the evaluation of heat moments in Eq.~\eqref{eq:main_result}. The heat operator (shaded in orange) defined in \eqref{eq:heat_operator} is applied $n$ times to the time-evolved state $\hat\rho(t)$ (denoted with circles) defined on $S$ and the extended bath $OA$. 
    }
    \label{fig:tn_diagram}
\end{figure}

\textcolor{newtext}{\textit{Calculation of heat moments.---}}\textcolor{newtext}{In case of macroscopic baths, direct simulation of the unitary evolution in Eq.~\eqref{eq:main_result} is a formidable task; however, it can be achieved efficiently using standard tensor-network wavefunction methods, e.g.,~the time evolving density matrix using orthogonal polynomials algorithm (TEDOPA)~\cite{prior_efficient_2010,Nuesseler2020}, or the multilayer multiconfiguration time-dependent Hartree (ML-MCTDH)~\cite{wang2003multilayer,wang2015multilayer} methods.} We choose the orthogonal polynomial approach (details in Appendix \ref{app:TEDOPA}) to represent thermodynamic operators and perform unitary time evolution of the global state using the time dependent variational principle (TDVP)~\cite{Haegeman2011}. A tensor-network diagram illustrating the evaluation of heat moments using Eq.~\eqref{eq:main_result} is shown in Fig.~\ref{fig:tn_diagram}.

\textcolor{newtext}{For a pure initial system state $|\psi_{\text{\tiny $S$}}(0)\rangle$, Eq.~\eqref{eq:main_result} becomes  $\langle Q ^n(t)\rangle$ $=  \langle\Psi_{\text{\tiny $G$}}(t)| \,\Tilde{\mathcal{Q}}^n \,|\Psi_{\text{\tiny $G$}}(t)\rangle$, where we prepare $|\Psi_{\text{\tiny $G$}}(t)\rangle$ $= \tilde{U}_{\text{\tiny $G$}}(t)|\psi_{\text{\tiny $S$}}(0)\rangle \otimes |\tilde{\Phi}\rangle$ using standard time propagation schemes mentioned above. After the chain mapping of $OA$ baths into $L_{\text{\tiny $B$}}$ bosonic sites each, the state $|\Psi_{\text{\tiny $G$}}(t)\rangle$ is defined on $L_{\text{\tiny $S$}}$ sites of the system and $2L_{\text{\tiny $B$}}$ of the baths. In case of a mixed initial state of the system, we can perform a spectral decomposition of $\hat\rho_{\text{\tiny $S$}}(0)=\sum_{k=1}^{d_\text{\tiny $S$}} p_k |k\rangle \langle k|$ and evaluate $\langle Q ^n(t)\rangle =  \sum_{k=1}^{d_\text{\tiny $S$}} p_k \langle\Psi_{\text{\tiny $G,k$}}(t)| \,\Tilde{\mathcal{Q}}^n \,|\Psi_{\text{\tiny $G,k$}}(t)\rangle$ with $|\Psi_{\text{\tiny $G,k$}}(t)\rangle = \tilde{U}_{\text{\tiny $G$}}(t)|k\rangle \otimes |\tilde{\Phi}\rangle$. If we denote by $C_n$ the computational complexity of evaluating the $n$-th heat moment of the pure system initial state (see Appendix \ref{app:complexity}), the same for mixed states using the decomposition above is approximately $d_\text{\tiny $S$}C_n$. Alternatively,  we can purify the mixed initial system state leading to a single global pure state defined on $2L_{\text{\tiny $B$}}+2L_{\text{\tiny $S$}}$ sites. Then the complexity is slightly higher than $C_n$, but generally better than the scaling with $d_{\text{\tiny $S$}}$ in the spectral decomposition approach. }

\textcolor{newtext}{The characteristic function from Eq.~\eqref{eq:simpl_chi_q} can be cast as a survival amplitude between the time-evolved state 
$|\Psi_{\text{\tiny $G$}}(t)\rangle$ and its $\lambda$–tilted counterpart $e^{i\lambda\tilde{\mathcal Q}}|\Psi_{\text{\tiny $G$}}(t)\rangle$. Since it is defined via a thermofield purification, this survival amplitude differs from the overlap that appears when measuring heat by an ancilla-assisted interferometric protocol~\cite{Goold2014}. Such ancilla-based protocols have previously been exploited in tensor-network simulations of characteristic functions for work~\cite{Johnson2016} or charge~\cite{Samajdar2024} fluctuations in many-body lattice systems.}

\textcolor{newtext}{\textit{Examples.---}}We implement our method to calculate the heat statistics in the paradigmatic example of a spin embedded in a bosonic environment, \textcolor{newtext}{considering both equilibrium and nonequilibrium settings of the same}. The equilibrium spin-boson model describes a two-level system (spin) interacting with a harmonic oscillator bath and stands as a cornerstone of quantum dissipation theory~\cite{weiss_quantum_2012}. It is described by the Hamiltonian $\hat{H} = \epsilon_0 \hat{S}_z +\Delta \hat{S}_x  + \sum_{\nu} \omega_{\nu} \hat{a}^{\dagger}_{\nu} \hat{a}_{\nu} + \hat{S}_x \otimes \sum\nolimits_{\nu} g_{\nu} ( \hat{a}_{\nu} + \hat{a}^{\dagger}_{\nu})$, with $\hat{S}_{x,z}$ the spin-1/2 angular momentum operators and $\hat{a}_{\nu}$ the bosonic annihilation operator for the bath mode $\nu$. The two-level system can be understood as representing the two lowest-energy states of a particle confined in a double-well potential. The energy splitting, $\epsilon_0$, corresponds to an asymmetry between the wells, while the tunneling amplitude, $\Delta$, quantifies the coupling between the two localized states due to tunneling through the central barrier. We assume the bath spectral density to be of Ohmic type, $J(\omega)= 2\alpha\omega\, \text{exp}(-\omega/\omega_C)$, with $\alpha$ a dimensionless coupling strength and $\omega_C$ a cutoff frequency. 

\begin{figure}
    \includegraphics[width=\linewidth]{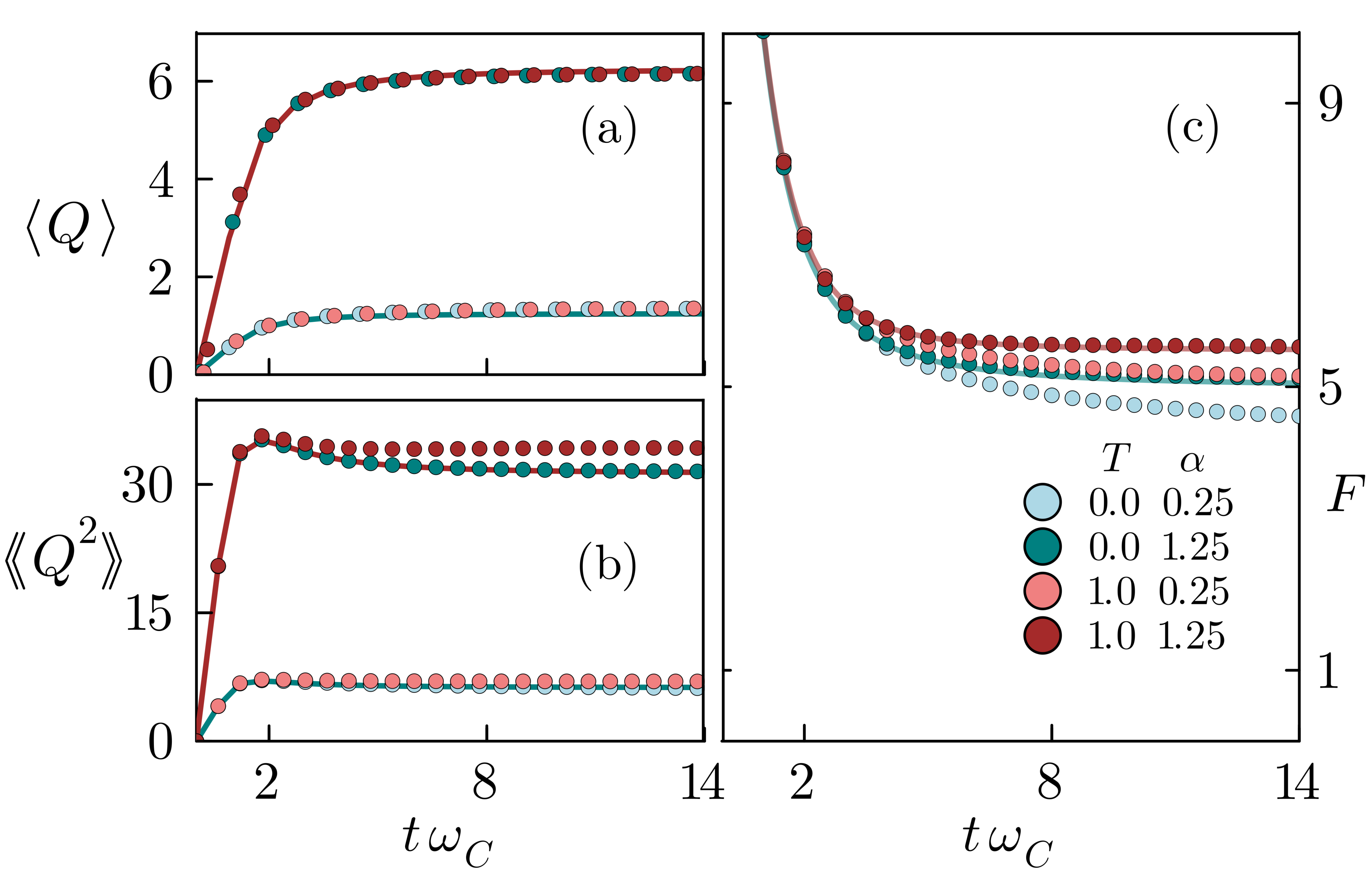}
    \caption{Evolution of (a)~mean, (b)~variance, and (c)~Fano factor of heat fluctuations in the spin-boson model ($\epsilon_0 = 1,\ \Delta = 0,\ \omega_C = 5 $). The initial state of the spin is the $|+\rangle$ eigenstate of $\hat{S}_x$. Curves corresponding to each $T$-$\alpha$ pair for the independent boson model (from benchmark results in Fig.\ \ref{fig:exact_model}) are shown as solid colored lines for reference.}
   \label{fig:fano}
\end{figure}
\textcolor{newtext}{As a benchmark, we first study the exactly solvable independent-boson limit ($\epsilon_0=0$, $\Delta=1$) of this model. As shown in Appendix \ref{app:indep-model}, }we see excellent agreement with the exact solution at all times for both zero and finite temperatures, and both weak ($\alpha < 1$) and strong ($\alpha\gtrsim 1$) bath couplings. In particular, we verify that the Fano factor, $F(t)=\langle \langle Q^2(t) \rangle \rangle/\langle Q(t) \rangle$, is independent of $\alpha$, in agreement with the exact solution which predicts that all heat cumulants are proportional to $\alpha$~\cite{popovic_quantum_2021}.  

In the nonintegrable regime ($\epsilon_0=1$, $\Delta=0$) of this model, there exists a quantum phase transition~\cite{leggett_dynamics_1987, bulla_numerical_2003} from the delocalized to the localized regime of the spin in the $\hat{S}_x$ eigenbasis when the bath coupling constant $\alpha$ exceeds the critical value $\alpha_c = 1+\mathcal{O}(\epsilon_0 / \omega_C)$. In Fig.\ \ref{fig:fano}, we examine the heat statistics for the coupling strengths on either side of this critical point. For both zero and finite temperatures, the mean and variance of the exchanged heat increases with coupling. For $\alpha > \alpha_c$, the heat statistics approach the predictions of the independent-boson model, since the spin dynamics is effectively frozen and the energetics are dominated by the system-bath interaction. For weaker couplings, the mean heat quickly equilibrates, while the variance exhibits two well separated timescales: a fast rise followed by a slow relaxation dictated by spin equilibration. These features are most clearly evident in the Fano factor, which depends significantly on both temperature and coupling strength, in contrast to the coupling-independent behavior of the independent-boson model in Fig.~\ref{fig:fano}.

\begin{figure}[t]
    \includegraphics[width=\linewidth]{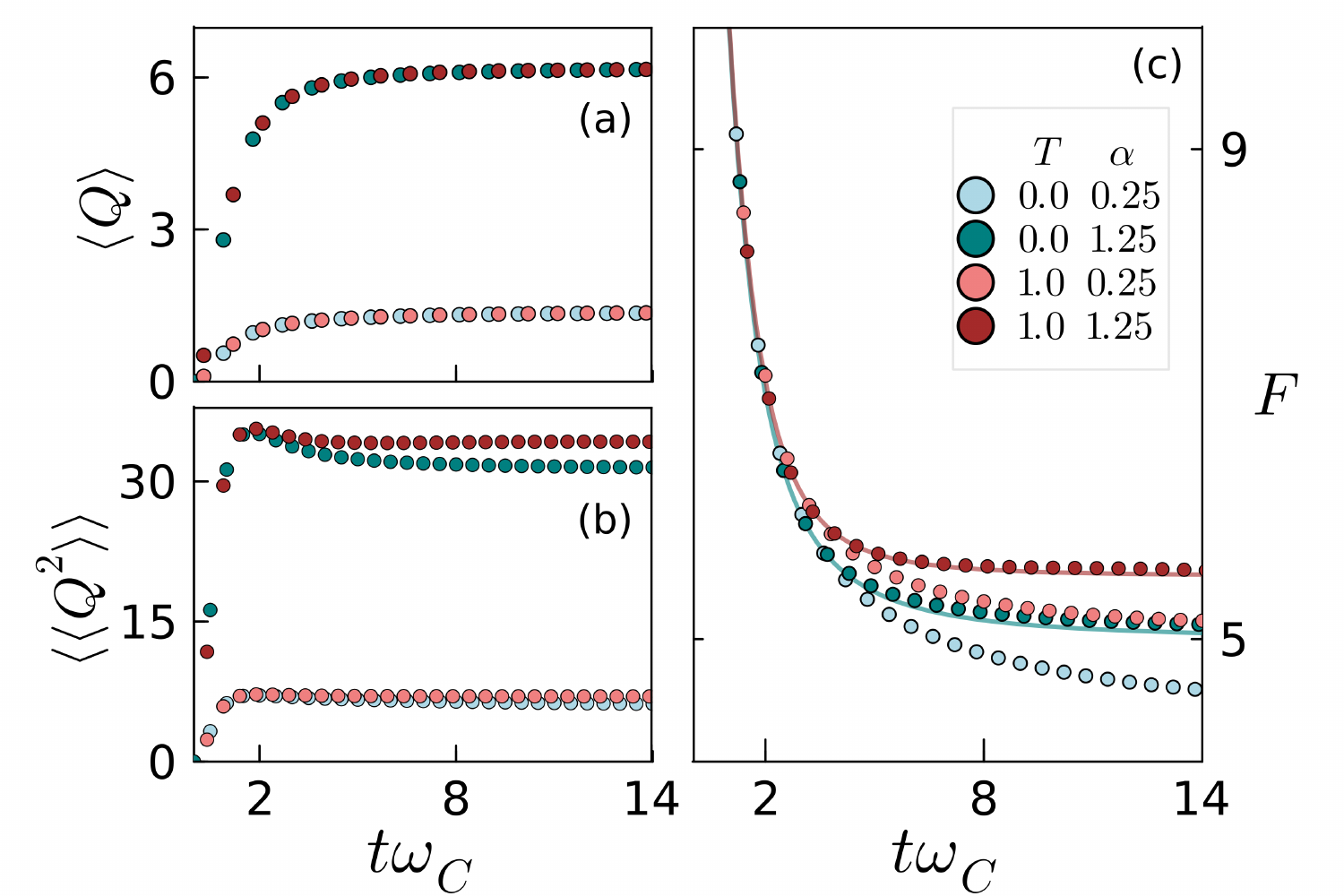}
    \caption{\textcolor{newtext}{Current fluctuations in the nonequilibrium spin–boson model. Time evolution of the (a) mean, (b) variance, and (c) Fano factor of the heat current for two bosonic baths with cutoff frequency $\omega_C$ at temperatures $T_1 = 1$ and $T_2 = 0$. We set $\epsilon_0 = 1$, $\Delta = 0$, and initialize the spin in the excited eigenstate of $\hat{S}_z$. The double-headed vertical arrow in (a) highlights the difference between the rectified currents (green and brown dots). A gray dashed line in (c) marks $F=1$ corresponding to Poissonian statistics. All vertical axes are plotted on a logarithmic scale.}}
   \label{fig:current_fano}
\end{figure}

\textcolor{newtext}{We further extend our analysis to the nonequilibrium spin-boson model, where the spin is coupled to two bosonic baths at different temperatures, $T_1=\epsilon_0$ and $T_2=0$ ($\epsilon_0=1$, $\Delta=0$). We compute the statistics of the heat current $J(t) = \delta_Q(t) /t$ by defining the integrated current $\delta_Q(t)=Q_2(t) - Q_1(t)$ flowing from bath 1 to bath 2 over time $t$. The scaled cumulants of the current are obtained as $\langle\langle J ^n(t)\rangle\rangle = \langle\langle \Tilde{\Delta}_Q ^n(t)\rangle\rangle /t$ with $\Tilde{\Delta}_Q(t)=\Tilde{\mathcal{Q}}_2(t)-\Tilde{\mathcal{Q}}_1(t)$, since $\Tilde{\mathcal{Q}}_1(t)$ commutes with $\Tilde{\mathcal{Q}}_2(t)$. Fig.~\ref{fig:current_fano} summarizes how the time-dependent mean and fluctuations of the current depend on the coupling strengths $(\alpha_1,\, \alpha_2)$ and cutoff frequencies $\omega_C$. For large coupling asymmetry, we observe thermal rectification, i.e.~the mean heat current for $\alpha_1 > \alpha_2$ is suppressed by an order of magnitude compared to when $\alpha_2 > \alpha_1$. However, the low-current configuration ($\alpha_1 \gg \alpha_2$) is characterized by significant current fluctuations and a very large Fano factor, indicating that this thermal diode blocks the current quite unreliably. Conversely, the reverse configuration ($\alpha_2 \gg \alpha_1$) exhibits a more regular current, with the long-time Fano factor approaching $F \approx 1$ for extreme coupling asymmetry, indicating nearly Poissonian (renewal-like) heat statistics. Crucially, this tendency in this conducting configuration survives down to $\omega_C=1$, where long bath memory ($\tau_B \sim \omega_C ^{-1}$) usually renders the evaluation of current fluctuations particularly challenging. While previous work has studied thermal rectification in spin-boson junctions at the level of average currents~\cite{segal2005spin, saito2013kondo,khandelwal2023characterizing}, our results demonstrate the importance of heat fluctuations for characterizing the performance of quantum devices in the strong-coupling regime.}

\textcolor{newtext}{\textit{Conclusions.---}}Heat exchange is a crucial aspect of open quantum dynamics, but its characterization in the strong-coupling regime has proved elusive, partly because of the two-time measurement implicit in its definition. We bypass this issue by defining the heat operator~\eqref{eq:heat_operator}, in terms of which the heat statistics can be extracted from dynamical pure state expectation values on an extended Hilbert space, as in Eq.~\eqref{eq:main_result}. \textcolor{newtext}{Our framework provides direct access to time-resolved, accurate heat-current statistics and complements previous analyses of the equilibrium and nonequilibrium spin-boson models beyond perturbative, high temperature, or steady state limits~\cite{nicolin2011quantum,liu2017energy, wang2017unifying,agarwalla2017energy}. In the equilibrium model, we uncover nontrivial dynamics of heat fluctuations at moderately weak coupling strengths, due to competition between the spin energy splitting and the system-bath interaction. In the nonequilibrium case, we characterize the reliability of a thermal diode and uncover a regime of current noise suppression, where the Fano factor approaches the Poissonian limit under extreme imbalance in coupling with the baths. The technique can also be extended to other quantities such as particle currents. Our work thus substantially expands the toolkit of quantum stochastic thermodynamics to encompass both steady state and transient dynamics in the strong-coupling regime.}
\vspace{3pt}
 
\textit{Data availability}.---The open-source code used in our simulations is based on the ITensor library~\cite{fishman_itensor_2022} and can be accessed at Ref.~\cite{github}.

\textit{Acknowledgments}.---We thank Saulo V. Moreira, Antonio Verd\'u, and Hari Kumar Yadalam for helpful discussions and suggestions regarding the manuscript. We warmly acknowledge Kesha for the use of her computational resources. This project is co-funded by the European Union (Quantum Flagship project ASPECTS, Grant Agreement No. 101080167) and UK Research and Innovation (UKRI). Views and opinions expressed are, however those of the authors only and do not necessarily reflect those of the European Union, Research Executive Agency or UKRI. Neither the European Union nor UKRI can be held responsible for them. M.T.M. is supported by a Royal Society University Research Fellowship. J.P. acknowledges support from grant TED2021-130578B-I00 and grant PID2021-124965NB-C21 funded by MICIU/AEI/10.13039/501100011033 and by the ``European Union NextGeneration EU/PRTR". 


\bibliographystyle{apsrev4-2} 
\bibliography{references}

@book{nielsenQuantumComputationQuantum2012,
  title = {Quantum {{Computation}} and {{Quantum Information}}: 10th {{Anniversary Edition}}},
  shorttitle = {Quantum {{Computation}} and {{Quantum Information}}},
  author = {Nielsen, Michael A. and Chuang, Isaac L.},
  year = 2012,
  month = jun,
  edition = {1},
  doi = {10.1017/CBO9780511976667},
  urldate = {2023-02-25},
  isbn = {978-1-107-00217-3 978-0-511-97666-7},
publisher={Cambridge university press}
}

@article{Colla2022,
  title     = {Open-system approach to nonequilibrium quantum thermodynamics at arbitrary coupling},
  author    = {Colla, Alessandra and Breuer, Heinz-Peter},
  journal   = {Phys. Rev. A},
  volume    = {105},
  issue     = {5},
  pages     = {052216},
  numpages  = {8},
  year      = {2022},
  month     = {May},
  publisher = {American Physical Society},
  doi       = {10.1103/PhysRevA.105.052216},
  url       = {https://link.aps.org/doi/10.1103/PhysRevA.105.052216}
}

@article{prior2013quantum,
  title     = {Quantum dynamics in photonic crystals},
  author    = {Prior, Javier and de Vega, Ines and Chin, Alex W and Huelga, Susana F and Plenio, Martin B},
  journal   = {Physical Review A—Atomic, Molecular, and Optical Physics},
  volume    = {87},
  number    = {1},
  pages     = {013428},
  year      = {2013},
  url       = {https://doi.org/10.1103/PhysRevA.87.013428},
  publisher = {APS}
}

@article{liu2017energy,
  title={Energy transfer in the nonequilibrium spin-boson model: From weak to strong coupling},
  author={Liu, Junjie and Xu, Hui and Li, Baowen and Wu, Changqin},
  journal={Physical Review E},
  volume={96},
  number={1},
  pages={012135},
  year={2017},
  publisher={APS},
url={https://journals.aps.org/pre/abstract/10.1103/PhysRevE.96.012135}
}

@article{khandelwal2023characterizing,
  title={Characterizing the performance of heat rectifiers},
  author={Khandelwal, Shishir and Perarnau-Llobet, Mart{\'\i} and Seah, Stella and Brunner, Nicolas and Haack, G{\'e}raldine},
  journal={Physical Review Research},
  volume={5},
  number={1},
  pages={013129},
  year={2023},
  publisher={APS},
url={https://journals.aps.org/prresearch/abstract/10.1103/PhysRevResearch.5.013129}
}

@article{agarwalla2017energy,
  title={Energy current and its statistics in the nonequilibrium spin-boson model: Majorana fermion representation},
  author={Agarwalla, Bijay Kumar and Segal, Dvira},
  journal={New Journal of Physics},
  volume={19},
  number={4},
  pages={043030},
  year={2017},
  publisher={IOP Publishing},
  url={https://iopscience.iop.org/article/10.1088/1367-2630/aa6657}
}

@article{wang2017unifying,
  title={Unifying quantum heat transfer in a nonequilibrium spin-boson model with full counting statistics},
  author={Wang, Chen and Ren, Jie and Cao, Jianshu},
  journal={Physical Review A},
  volume={95},
  number={2},
  pages={023610},
  year={2017},
  publisher={APS},
  url={https://journals.aps.org/pra/abstract/10.1103/PhysRevA.95.023610}
}

@article{aurell2018characteristic,
  title={Characteristic functions of quantum heat with baths at different temperatures},
  author={Aurell, Erik},
  journal={Physical Review E},
  volume={97},
  number={6},
  pages={062117},
  year={2018},
  publisher={APS},
url = {https://journals.aps.org/pre/abstract/10.1103/PhysRevE.97.062117}
}

@article{nicolin2011quantum,
  title={Quantum fluctuation theorem for heat exchange in the strong coupling regime},
  author={Nicolin, Lena and Segal, Dvira},
  journal={Physical Review B—Condensed Matter and Materials Physics},
  volume={84},
  number={16},
  pages={161414},
  year={2011},
  publisher={APS},
  url={https://journals.aps.org/prb/abstract/10.1103/PhysRevB.84.161414}
}

@article{wang2003multilayer,
  title     = {Multilayer formulation of the multiconfiguration time-dependent Hartree theory},
  author    = {Wang, Haobin and Thoss, Michael},
  journal   = {The Journal of chemical physics},
  volume    = {119},
  number    = {3},
  pages     = {1289--1299},
  year      = {2003},
  publisher = {American Institute of Physics},
url={https://pubs.aip.org/aip/jcp/article-abstract/119/3/1289/186240/Multilayer-formulation-of-the-multiconfiguration?redirectedFrom=fulltext}
}

@article{Strasberg2021,
  title     = {First and Second Law of Quantum Thermodynamics: A Consistent Derivation Based on a Microscopic Definition of Entropy},
  author    = {Strasberg, Philipp and Winter, Andreas},
  journal   = {PRX Quantum},
  volume    = {2},
  issue     = {3},
  pages     = {030202},
  numpages  = {26},
  year      = {2021},
  month     = {Aug},
  publisher = {American Physical Society},
  doi       = {10.1103/PRXQuantum.2.030202},
  url       = {https://link.aps.org/doi/10.1103/PRXQuantum.2.030202}
}

@article{Samajdar2024,
  title     = {Quantum Turnstiles for Robust Measurement of Full Counting Statistics},
  author    = {Samajdar, Rhine and McCulloch, Ewan and Khemani, Vedika and Vasseur, Romain and Gopalakrishnan, Sarang},
  journal   = {Phys. Rev. Lett.},
  volume    = {133},
  issue     = {24},
  pages     = {240403},
  numpages  = {8},
  year      = {2024},
  month     = {Dec},
  publisher = {American Physical Society},
  doi       = {10.1103/PhysRevLett.133.240403},
  url       = {https://link.aps.org/doi/10.1103/PhysRevLett.133.240403}
}

@article{Yadalam2022,
  title     = {Counting statistics of energy transport across squeezed thermal reservoirs},
  author    = {Yadalam, Hari Kumar and Agarwalla, Bijay Kumar and Harbola, Upendra},
  journal   = {Phys. Rev. A},
  volume    = {105},
  issue     = {6},
  pages     = {062219},
  numpages  = {14},
  year      = {2022},
  month     = {Jun},
  publisher = {American Physical Society},
  doi       = {10.1103/PhysRevA.105.062219},
  url       = {https://link.aps.org/doi/10.1103/PhysRevA.105.062219}
}

@article{Wang2014,
  title     = {Nonequilibrium {{Green}}'s Function Method for Quantum Thermal Transport},
  author    = {Wang, Jian-Sheng and Agarwalla, Bijay Kumar and Li, Huanan and Thingna, Juzar},
  year      = {2014},
  month     = dec,
  journal   = {Frontiers of Physics},
  volume    = {9},
  number    = {6},
  pages     = {673--697},
  issn      = {2095-0462, 2095-0470},
  doi       = {10.1007/s11467-013-0340-x},
  urldate   = {2025-04-13},
  copyright = {http://www.springer.com/tdm},
  langid    = {english}
}

@article{Goold2014,
  title     = {Measuring the heat exchange of a quantum process},
  author    = {Goold, John and Poschinger, Ulrich and Modi, Kavan},
  journal   = {Phys. Rev. E},
  volume    = {90},
  issue     = {2},
  pages     = {020101},
  numpages  = {4},
  year      = {2014},
  month     = {Aug},
  publisher = {American Physical Society},
  doi       = {10.1103/PhysRevE.90.020101},
  url       = {https://link.aps.org/doi/10.1103/PhysRevE.90.020101}
}

@misc{github,
  title        = {{ASPECTS} code repository},
  howpublished = {\url{https://github.com/aspects-quantum/TFlucn_tedopa.git}},
year = {2024}
}

@article{Haegeman2011,
  title     = {Time-Dependent Variational Principle for Quantum Lattices},
  author    = {Haegeman, Jutho and Cirac, J. Ignacio and Osborne, Tobias J. and Pi\ifmmode \check{z}\else \v{z}\fi{}orn, Iztok and Verschelde, Henri and Verstraete, Frank},
  journal   = {Phys. Rev. Lett.},
  volume    = {107},
  issue     = {7},
  pages     = {070601},
  numpages  = {5},
  year      = {2011},
  month     = {Aug},
  publisher = {American Physical Society},
  doi       = {10.1103/PhysRevLett.107.070601},
  url       = {https://link.aps.org/doi/10.1103/PhysRevLett.107.070601}
}

@article{Tamascelli2018,
  title     = {Nonperturbative Treatment of non-Markovian Dynamics of Open Quantum Systems},
  author    = {Tamascelli, D. and Smirne, A. and Huelga, S. F. and Plenio, M. B.},
  journal   = {Phys. Rev. Lett.},
  volume    = {120},
  issue     = {3},
  pages     = {030402},
  numpages  = {6},
  year      = {2018},
  month     = {Jan},
  publisher = {American Physical Society},
  doi       = {10.1103/PhysRevLett.120.030402},
  url       = {https://link.aps.org/doi/10.1103/PhysRevLett.120.030402}
}

@article{Nuesseler2020,
  title     = {Efficient simulation of open quantum systems coupled to a fermionic bath},
  author    = {N\"u\ss{}eler, Alexander and Dhand, Ish and Huelga, Susana F. and Plenio, Martin B.},
  journal   = {Phys. Rev. B},
  volume    = {101},
  issue     = {15},
  pages     = {155134},
  numpages  = {20},
  year      = {2020},
  month     = {Apr},
  publisher = {American Physical Society},
  doi       = {10.1103/PhysRevB.101.155134},
  url       = {https://link.aps.org/doi/10.1103/PhysRevB.101.155134}
}

@article{Tamascelli2019,
  title     = {Efficient Simulation of Finite-Temperature Open Quantum Systems},
  author    = {Tamascelli, D. and Smirne, A. and Lim, J. and Huelga, S. F. and Plenio, M. B.},
  journal   = {Phys. Rev. Lett.},
  volume    = {123},
  issue     = {9},
  pages     = {090402},
  numpages  = {6},
  year      = {2019},
  month     = {Aug},
  publisher = {American Physical Society},
  doi       = {10.1103/PhysRevLett.123.090402},
  url       = {https://link.aps.org/doi/10.1103/PhysRevLett.123.090402}
}

@article{Purkayastha2021,
  title     = {Periodically refreshed baths to simulate open quantum many-body dynamics},
  author    = {Purkayastha, Archak and Guarnieri, Giacomo and Campbell, Steve and Prior, Javier and Goold, John},
  journal   = {Phys. Rev. B},
  volume    = {104},
  issue     = {4},
  pages     = {045417},
  numpages  = {16},
  year      = {2021},
  month     = {Jul},
  publisher = {American Physical Society},
  doi       = {10.1103/PhysRevB.104.045417},
  url       = {https://link.aps.org/doi/10.1103/PhysRevB.104.045417}
}

@article{Somoza2019,
  title     = {Dissipation-Assisted Matrix Product Factorization},
  author    = {Somoza, Alejandro D. and Marty, Oliver and Lim, James and Huelga, Susana F. and Plenio, Martin B.},
  journal   = {Phys. Rev. Lett.},
  volume    = {123},
  issue     = {10},
  pages     = {100502},
  numpages  = {7},
  year      = {2019},
  month     = {Sep},
  publisher = {American Physical Society},
  doi       = {10.1103/PhysRevLett.123.100502},
  url       = {https://link.aps.org/doi/10.1103/PhysRevLett.123.100502}
}

@article{Thoeniss2023,
  title     = {Efficient method for quantum impurity problems out of equilibrium},
  author    = {Thoenniss, Julian and Sonner, Michael and Lerose, Alessio and Abanin, Dmitry A.},
  journal   = {Phys. Rev. B},
  volume    = {107},
  issue     = {20},
  pages     = {L201115},
  numpages  = {6},
  year      = {2023},
  month     = {May},
  publisher = {American Physical Society},
  doi       = {10.1103/PhysRevB.107.L201115},
  url       = {https://link.aps.org/doi/10.1103/PhysRevB.107.L201115}
}

@article{Javier09,
  title     = {Density Matrix Renormalization Group in the Heisenberg Picture},
  author    = {Hartmann, Michael J. and Prior, Javier and Clark, Stephen R. and Plenio, Martin B.},
  journal   = {Phys. Rev. Lett.},
  volume    = {102},
  issue     = {5},
  pages     = {057202},
  numpages  = {4},
  year      = {2009},
  month     = {Feb},
  publisher = {American Physical Society},
  doi       = {10.1103/PhysRevLett.102.057202},
  url       = {https://link.aps.org/doi/10.1103/PhysRevLett.102.057202}
}

@article{Carrega2016,
  title     = {Energy Exchange in Driven Open Quantum Systems at Strong Coupling},
  author    = {Carrega, Matteo and Solinas, Paolo and Sassetti, Maura and Weiss, Ulrich},
  journal   = {Phys. Rev. Lett.},
  volume    = {116},
  issue     = {24},
  pages     = {240403},
  numpages  = {5},
  year      = {2016},
  month     = {Jun},
  publisher = {American Physical Society},
  doi       = {10.1103/PhysRevLett.116.240403},
  url       = {https://link.aps.org/doi/10.1103/PhysRevLett.116.240403}
}

@article{Song2017,
  title     = {Hierarchical equations of motion method applied to nonequilibrium heat transport in model molecular junctions: Transient heat current and high-order moments of the current operator},
  author    = {Song, Linze and Shi, Qiang},
  journal   = {Phys. Rev. B},
  volume    = {95},
  issue     = {6},
  pages     = {064308},
  numpages  = {10},
  year      = {2017},
  month     = {Feb},
  publisher = {American Physical Society},
  doi       = {10.1103/PhysRevB.95.064308},
  url       = {https://link.aps.org/doi/10.1103/PhysRevB.95.064308}
}

@article{Shubrook2025,
  title     = {Non-{{Markovian}} Quantum Heat Statistics with the Reaction Coordinate Mapping},
  author    = {Shubrook, Mike and {Iles-Smith}, Jake and Nazir, Ahsan},
  year      = {2025},
  month     = apr,
  journal   = {Quantum Science and Technology},
  volume    = {10},
  number    = {2},
  pages     = {025063},
  publisher = {IOP Publishing},
  issn      = {2058-9565},
  doi       = {10.1088/2058-9565/adc6b7},
  urldate   = {2025-04-30},
  langid    = {english}
}

@article{Segal2016,
  title     = {Vibrational {{Heat Transport}} in {{Molecular Junctions}}},
  author    = {Segal, Dvira and Agarwalla, Bijay Kumar},
  year      = {2016},
  month     = may,
  journal   = {Annual Review of Physical Chemistry},
  volume    = {67},
  number    = {Volume 67, 2016},
  pages     = {185--209},
  publisher = {Annual Reviews},
  issn      = {0066-426X, 1545-1593},
  doi       = {10.1146/annurev-physchem-040215-112103},
  urldate   = {2025-04-06},
  langid    = {english}
}

@article{Wojtowicz2020,
  title     = {Open-system tensor networks and Kramers' crossover for quantum transport},
  author    = {W\'ojtowicz, Gabriela and Elenewski, Justin E. and Rams, Marek M. and Zwolak, Michael},
  journal   = {Phys. Rev. A},
  volume    = {101},
  issue     = {5},
  pages     = {050301},
  numpages  = {7},
  year      = {2020},
  month     = {May},
  publisher = {American Physical Society},
  doi       = {10.1103/PhysRevA.101.050301},
  url       = {https://link.aps.org/doi/10.1103/PhysRevA.101.050301}
}

@article{Lotem2020,
  title     = {Renormalized Lindblad driving: A numerically exact nonequilibrium quantum impurity solver},
  author    = {Lotem, Matan and Weichselbaum, Andreas and von Delft, Jan and Goldstein, Moshe},
  journal   = {Phys. Rev. Res.},
  volume    = {2},
  issue     = {4},
  pages     = {043052},
  numpages  = {17},
  year      = {2020},
  month     = {Oct},
  publisher = {American Physical Society},
  doi       = {10.1103/PhysRevResearch.2.043052},
  url       = {https://link.aps.org/doi/10.1103/PhysRevResearch.2.043052}
}

@article{Bettmann2024,
  title={Quantum stochastic thermodynamics in the mesoscopic-leads formulation},
  author={Bettmann, Laetitia P and Kewming, Michael J and Landi, Gabriel T and Goold, John and Mitchison, Mark T},
  journal={Physical Review E},
  volume={112},
  number={1},
  pages={014105},
  year={2025},
  publisher={APS},
url={https://journals.aps.org/pre/abstract/10.1103/3msj-9qgb}
}

@article{Brenes2023,
  title     = {Particle current statistics in driven mesoscale conductors},
  author    = {Brenes, Marlon and Guarnieri, Giacomo and Purkayastha, Archak and Eisert, Jens and Segal, Dvira and Landi, Gabriel},
  journal   = {Phys. Rev. B},
  volume    = {108},
  issue     = {8},
  pages     = {L081119},
  numpages  = {6},
  year      = {2023},
  month     = {Aug},
  publisher = {American Physical Society},
  doi       = {10.1103/PhysRevB.108.L081119},
  url       = {https://link.aps.org/doi/10.1103/PhysRevB.108.L081119}
}

@article{Cirio2024,
  title     = {Pseudofermion method for the exact description of fermionic environments: From single-molecule electronics to the Kondo resonance},
  author    = {Cirio, Mauro and Lambert, Neill and Liang, Pengfei and Kuo, Po-Chen and Chen, Yueh-Nan and Menczel, Paul and Funo, Ken and Nori, Franco},
  journal   = {Phys. Rev. Res.},
  volume    = {5},
  issue     = {3},
  pages     = {033011},
  numpages  = {16},
  year      = {2023},
  month     = {Jul},
  publisher = {American Physical Society},
  doi       = {10.1103/PhysRevResearch.5.033011},
  url       = {https://link.aps.org/doi/10.1103/PhysRevResearch.5.033011}
}

@article{Menczel2024,
  title     = {Non-Hermitian pseudomodes for strongly coupled open quantum systems: Unravelings, correlations, and thermodynamics},
  author    = {Menczel, Paul and Funo, Ken and Cirio, Mauro and Lambert, Neill and Nori, Franco},
  journal   = {Phys. Rev. Res.},
  volume    = {6},
  issue     = {3},
  pages     = {033237},
  numpages  = {25},
  year      = {2024},
  month     = {Sep},
  publisher = {American Physical Society},
  doi       = {10.1103/PhysRevResearch.6.033237},
  url       = {https://link.aps.org/doi/10.1103/PhysRevResearch.6.033237}
}

@article{Brenes2020,
  title     = {Tensor-Network Method to Simulate Strongly Interacting Quantum Thermal Machines},
  author    = {Brenes, Marlon and Mendoza-Arenas, Juan Jos\'e and Purkayastha, Archak and Mitchison, Mark T. and Clark, Stephen R. and Goold, John},
  journal   = {Phys. Rev. X},
  volume    = {10},
  issue     = {3},
  pages     = {031040},
  numpages  = {29},
  year      = {2020},
  month     = {Aug},
  publisher = {American Physical Society},
  doi       = {10.1103/PhysRevX.10.031040},
  url       = {https://link.aps.org/doi/10.1103/PhysRevX.10.031040}
}

@article{Valli2024,
  title={Efficient computation of cumulant evolution and full counting statistics: application to infinite temperature quantum spin chains},
  author={Valli, Angelo and Moca, C{\u{a}}t{\u{a}}lin Pa{\c{s}}cu and Werner, Mikl{\'o}s Antal and Kormos, M{\'a}rton and Krajnik, {\v{Z}}iga and Prosen, Toma{\v{z}} and Zar{\'a}nd, Gergely},
  journal={Physical Review Letters},
  volume={135},
  number={10},
  pages={100401},
  year={2025},
  publisher={APS},
url={https://journals.aps.org/prl/abstract/10.1103/f3c4-n21z}
}

@article{Josefsson2019,
  title     = {Optimal power and efficiency of single quantum dot heat engines: Theory and experiment},
  author    = {Josefsson, Martin and Svilans, Artis and Linke, Heiner and Leijnse, Martin},
  journal   = {Phys. Rev. B},
  volume    = {99},
  issue     = {23},
  pages     = {235432},
  numpages  = {11},
  year      = {2019},
  month     = {Jun},
  publisher = {American Physical Society},
  doi       = {10.1103/PhysRevB.99.235432},
  url       = {https://link.aps.org/doi/10.1103/PhysRevB.99.235432}
}

@article{Trushechkin2022,
  title   = {Open Quantum System Dynamics and the Mean Force {{Gibbs}} State},
  author  = {Trushechkin, A. S. and Merkli, M. and Cresser, J. D. and Anders, J.},
  year    = {2022},
  month   = mar,
  journal = {AVS Quantum Science},
  volume  = {4},
  number  = {1},
  pages   = {012301},
  issn    = {2639-0213},
  doi     = {10.1116/5.0073853}
}

@article{breuer2016,
  title     = {Colloquium: Non-Markovian dynamics in open quantum systems},
  author    = {Breuer, Heinz-Peter and Laine, Elsi-Mari and Piilo, Jyrki and Vacchini, Bassano},
  journal   = {Rev. Mod. Phys.},
  volume    = {88},
  issue     = {2},
  pages     = {021002},
  numpages  = {24},
  year      = {2016},
  month     = {Apr},
  publisher = {American Physical Society},
  doi       = {10.1103/RevModPhys.88.021002},
  url       = {https://link.aps.org/doi/10.1103/RevModPhys.88.021002}
}

@article{Miller2020,
  title     = {Quantum Fluctuations Hinder Finite-Time Information Erasure near the Landauer Limit},
  author    = {Miller, Harry J. D. and Guarnieri, Giacomo and Mitchison, Mark T. and Goold, John},
  journal   = {Phys. Rev. Lett.},
  volume    = {125},
  issue     = {16},
  pages     = {160602},
  numpages  = {6},
  year      = {2020},
  month     = {Oct},
  publisher = {American Physical Society},
  doi       = {10.1103/PhysRevLett.125.160602},
  url       = {https://link.aps.org/doi/10.1103/PhysRevLett.125.160602}
}

@article{Spiecker2023,
  title     = {Two-Level System Hyperpolarization Using a Quantum {{Szilard}} Engine},
  author    = {Spiecker, Martin and Paluch, Patrick and Gosling, Nicolas and Drucker, Niv and Matityahu, Shlomi and Gusenkova, Daria and G{\"u}nzler, Simon and Rieger, Dennis and Takmakov, Ivan and Valenti, Francesco and Winkel, Patrick and Gebauer, Richard and Sander, Oliver and Catelani, Gianluigi and Shnirman, Alexander and Ustinov, Alexey V. and Wernsdorfer, Wolfgang and Cohen, Yonatan and Pop, Ioan M.},
  year      = {2023},
  month     = sep,
  journal   = {Nature Physics},
  volume    = {19},
  number    = {9},
  pages     = {1320--1325},
  publisher = {Nature Publishing Group},
  issn      = {1745-2481},
  doi       = {10.1038/s41567-023-02082-8},
  copyright = {2023 The Author(s), under exclusive licence to Springer Nature Limited},
  langid    = {english}
}

@article{AntonioMarinGuzman2024,
  title      = {Key Issues Review: Useful Autonomous Quantum Machines},
  shorttitle = {Key Issues Review},
  author     = {Antonio Mar{\'i}n Guzm{\'a}n, Jos{\'e} and Erker, Paul and Gasparinetti, Simone and Huber, Marcus and Yunger Halpern, Nicole},
  year       = {2024},
  month      = nov,
  journal    = {Rep. Prog. Phys.},
  volume     = {87},
  number     = {12},
  pages      = {122001},
  publisher  = {IOP Publishing},
  issn       = {0034-4885},
  doi        = {10.1088/1361-6633/ad8803},
  langid     = {english}
}

@article{Landi2021,
  title     = {Irreversible entropy production: From classical to quantum},
  author    = {Landi, Gabriel T. and Paternostro, Mauro},
  journal   = {Rev. Mod. Phys.},
  volume    = {93},
  issue     = {3},
  pages     = {035008},
  numpages  = {58},
  year      = {2021},
  month     = {Sep},
  publisher = {American Physical Society},
  doi       = {10.1103/RevModPhys.93.035008},
  url       = {https://link.aps.org/doi/10.1103/RevModPhys.93.035008}
}

@article{Popovic2023,
  title     = {Thermodynamics of Decoherence},
  author    = {Popovic, Maria and Mitchison, Mark T. and Goold, John},
  year      = {2023},
  month     = apr,
  journal   = {Proc. Roy. Soc. A},
  volume    = {479},
  number    = {2272},
  pages     = {20230040},
  publisher = {Royal Society},
  doi       = {10.1098/rspa.2023.0040}
}

@article{Kerremans2022,
  title     = {{Probabilistically violating the first law of thermodynamics in a quantum heat engine}},
  author    = {Timo Kerremans and Peter Samuelsson and Patrick P. Potts},
  journal   = {SciPost Phys.},
  volume    = {12},
  pages     = {168},
  year      = {2022},
  publisher = {SciPost},
  doi       = {10.21468/SciPostPhys.12.5.168},
  url       = {https://scipost.org/10.21468/SciPostPhys.12.5.168}
}

@article{Johnson2016,
  title     = {Thermometry of ultracold atoms via nonequilibrium work distributions},
  author    = {Johnson, T. H. and Cosco, F. and Mitchison, M. T. and Jaksch, D. and Clark, S. R.},
  journal   = {Phys. Rev. A},
  volume    = {93},
  issue     = {5},
  pages     = {053619},
  numpages  = {6},
  year      = {2016},
  month     = {May},
  publisher = {American Physical Society},
  doi       = {10.1103/PhysRevA.93.053619},
  url       = {https://link.aps.org/doi/10.1103/PhysRevA.93.053619}
}

@article{karrasch_finite-temperature_2012,
  title     = {Finite-{Temperature} {Dynamical} {Density} {Matrix} {Renormalization} {Group} and the {Drude} {Weight} of {Spin}- 1 / 2 {Chains}},
  volume    = {108},
  copyright = {http://link.aps.org/licenses/aps-default-license},
  issn      = {0031-9007, 1079-7114},
  url       = {https://link.aps.org/doi/10.1103/PhysRevLett.108.227206},
  doi       = {10.1103/PhysRevLett.108.227206},
  language  = {en},
  number    = {22},
  urldate   = {2024-06-24},
  journal   = {Phys. Rev. Lett.},
  author    = {Karrasch, C. and Bardarson, J. H. and Moore, J. E.},
  month     = may,
  year      = {2012},
  pages     = {227206}
}

@misc{breuer_theory_2007,
  edition   = {1},
  title     = {The {Theory} of {Open} {Quantum} {Systems}},
  isbn      = {978-0-19-921390-0 978-0-19-170634-9},
  url       = {\url{https://academic.oup.com/book/27757}},
  urldate   = {2024-06-17},
  publisher = {Oxford University PressOxford},
  author    = {Breuer, Heinz-Peter and Petruccione, Francesco},
  year      = {2007},
  doi       = {10.1093/acprof:oso/9780199213900.001.0001}
}

@book{weiss_quantum_2012,
  author    = {Weiss, Ulrich},
  title     = {An Introduction to Stochastic Thermodynamics From Basic to Advanced},
  url       = {https://doi.org/10.1142/8334},
  year      = {2012},
  publisher = {WORLD SCIENTIFIC},
  note      = {}
}

@article{de_vega_thermofield-based_2015,
  title    = {Thermofield-based chain-mapping approach for open quantum systems},
  volume   = {92},
  url      = {https://link.aps.org/doi/10.1103/PhysRevA.92.052116},
  doi      = {10.1103/PhysRevA.92.052116},
  number   = {5},
  urldate  = {2024-06-17},
  journal  = {Phys. Rev. A},
  author   = {de Vega, Inés and Bañuls, Mari-Carmen},
  month    = nov,
  year     = {2015},
  pages    = {052116}
}

@article{esposito_nonequilibrium_2009,
  title    = {Nonequilibrium fluctuations, fluctuation theorems, and counting statistics in quantum systems},
  volume   = {81},
  url      = {https://link.aps.org/doi/10.1103/RevModPhys.81.1665},
  doi      = {10.1103/RevModPhys.81.1665},
  number   = {4},
  urldate  = {2024-06-16},
  journal  = {Rev. Mod. Phys.},
  author   = {Esposito, Massimiliano and Harbola, Upendra and Mukamel, Shaul},
  month    = dec,
  year     = {2009},
  pages    = {1665--1702}
}

@article{prior_efficient_2010,
  title    = {Efficient {Simulation} of {Strong} {System}-{Environment} {Interactions}},
  volume   = {105},
  url      = {https://link.aps.org/doi/10.1103/PhysRevLett.105.050404},
  doi      = {10.1103/PhysRevLett.105.050404},
  number   = {5},
  urldate  = {2024-06-16},
  journal  = {Phys. Rev. Lett.},
  author   = {Prior, Javier and Chin, Alex W. and Huelga, Susana F. and Plenio, Martin B.},
  month    = jul,
  year     = {2010},
  pages    = {050404}
}

@article{talkner_colloquium_2020,
  title      = {\textit{{Colloquium}} : {Statistical} mechanics and thermodynamics at strong coupling: {Quantum} and classical},
  volume     = {92},
  issn       = {0034-6861, 1539-0756},
  shorttitle = {\textit{{Colloquium}}},
  url        = {https://link.aps.org/doi/10.1103/RevModPhys.92.041002},
  doi        = {10.1103/RevModPhys.92.041002},
  language   = {en},
  number     = {4},
  urldate    = {2024-06-06},
  journal    = {Rev. Mod. Phys.},
  author     = {Talkner, Peter and Hänggi, Peter},
  month      = oct,
  year       = {2020},
  pages      = {041002}
}

@article{popovic_quantum_2021,
  title    = {Quantum {Heat} {Statistics} with {Time}-{Evolving} {Matrix} {Product} {Operators}},
  volume   = {2},
  url      = {https://link.aps.org/doi/10.1103/PRXQuantum.2.020338},
  doi      = {10.1103/PRXQuantum.2.020338},
  number   = {2},
  urldate  = {2024-06-06},
  journal  = {PRX Quantum},
  author   = {Popovic, Maria and Mitchison, Mark T. and Strathearn, Aidan and Lovett, Brendon W. and Goold, John and Eastham, Paul R.},
  month    = jun,
  year     = {2021},
  pages    = {020338}
}

@article{kilgour_path-integral_2019,
  title      = {Path-integral methodology and simulations of quantum thermal transport: {Full} counting statistics approach},
  volume     = {150},
  issn       = {0021-9606, 1089-7690},
  shorttitle = {Path-integral methodology and simulations of quantum thermal transport},
  url        = {https://pubs.aip.org/jcp/article/150/8/084111/197467/Path-integral-methodology-and-simulations-of},
  doi        = {10.1063/1.5084949},
  language   = {en},
  number     = {8},
  urldate    = {2024-05-29},
  journal    = {The Journal of Chemical Physics},
  author     = {Kilgour, Michael and Agarwalla, Bijay Kumar and Segal, Dvira},
  month      = feb,
  year       = {2019},
  pages      = {084111}
}

@article{aamir_thermally_2025,
  title     = {Thermally driven quantum refrigerator autonomously resets a superconducting qubit},
  copyright = {2025 The Author(s)},
  issn      = {1745-2481},
  url       = {https://www.nature.com/articles/s41567-024-02708-5},
  doi       = {10.1038/s41567-024-02708-5},
  language  = {en},
  urldate   = {2025-02-12},
  journal   = {Nature Physics},
  author    = {Aamir, Mohammed Ali and Jamet Suria, Paul and Marín Guzmán, José Antonio and Castillo-Moreno, Claudia and Epstein, Jeffrey M. and Yunger Halpern, Nicole and Gasparinetti, Simone},
  month     = jan,
  year      = {2025},
  pages     = {1--6}
}

@article{ptaszynski_coherence-enhanced_2018,
  title    = {Coherence-enhanced constancy of a quantum thermoelectric generator},
  volume   = {98},
  url      = {https://link.aps.org/doi/10.1103/PhysRevB.98.085425},
  doi      = {10.1103/PhysRevB.98.085425},
  number   = {8},
  urldate  = {2023-05-29},
  journal  = {Phys. Rev. B},
  author   = {Ptaszyński, Krzysztof},
  month    = aug,
  year     = {2018},
  pages    = {085425}
}

@article{agarwalla_assessing_2018,
  title    = {Assessing the validity of the thermodynamic uncertainty relation in quantum systems},
  volume   = {98},
  url      = {https://link.aps.org/doi/10.1103/PhysRevB.98.155438},
  doi      = {10.1103/PhysRevB.98.155438},
  number   = {15},
  urldate  = {2023-05-29},
  journal  = {Phys. Rev. B},
  author   = {Agarwalla, Bijay Kumar and Segal, Dvira},
  month    = oct,
  year     = {2018},
  pages    = {155438}
}

@article{guarnieri_thermodynamics_2019,
  title    = {Thermodynamics of precision in quantum nonequilibrium steady states},
  volume   = {1},
  url      = {https://link.aps.org/doi/10.1103/PhysRevResearch.1.033021},
  doi      = {10.1103/PhysRevResearch.1.033021},
  number   = {3},
  urldate  = {2023-05-29},
  journal  = {Phys. Rev. Research},
  author   = {Guarnieri, Giacomo and Landi, Gabriel T. and Clark, Stephen R. and Goold, John},
  month    = oct,
  year     = {2019},
  pages    = {033021}
}

@article{kalaee_violating_2021,
  title     = {Violating the thermodynamic uncertainty relation in the three-level maser},
  author    = {Kalaee, Alex Arash Sand and Wacker, Andreas and Potts, Patrick P.},
  journal   = {Phys. Rev. E},
  volume    = {104},
  issue     = {1},
  pages     = {L012103},
  numpages  = {6},
  year      = {2021},
  month     = {Jul},
  publisher = {American Physical Society},
  doi       = {10.1103/PhysRevE.104.L012103},
  url       = {https://link.aps.org/doi/10.1103/PhysRevE.104.L012103}
}

@article{rignon-bret_thermodynamics_2021,
  title    = {Thermodynamics of precision in quantum nanomachines},
  volume   = {103},
  url      = {https://link.aps.org/doi/10.1103/PhysRevE.103.012133},
  doi      = {10.1103/PhysRevE.103.012133},
  number   = {1},
  urldate  = {2023-05-29},
  journal  = {Phys. Rev. E},
  author   = {Rignon-Bret, Antoine and Guarnieri, Giacomo and Goold, John and Mitchison, Mark T.},
  month    = jan,
  year     = {2021},
  pages    = {012133}
}

@article{brandner_thermodynamic_2018,
  title    = {Thermodynamic {Bounds} on {Precision} in {Ballistic} {Multiterminal} {Transport}},
  volume   = {120},
  url      = {https://link.aps.org/doi/10.1103/PhysRevLett.120.090601},
  doi      = {10.1103/PhysRevLett.120.090601},
  number   = {9},
  urldate  = {2023-09-17},
  journal  = {Phys. Rev. Lett.},
  author   = {Brandner, Kay and Hanazato, Taro and Saito, Keiji},
  month    = mar,
  year     = {2018},
  pages    = {090601}
}

@article{rivas_quantum_2014,
  title      = {Quantum non-{Markovianity}: characterization, quantification and detection},
  volume     = {77},
  issn       = {0034-4885, 1361-6633},
  shorttitle = {Quantum non-{Markovianity}},
  url        = {https://iopscience.iop.org/article/10.1088/0034-4885/77/9/094001},
  doi        = {10.1088/0034-4885/77/9/094001},
  number     = {9},
  urldate    = {2023-03-06},
  journal    = {Rep. Prog. Phys.},
  author     = {Rivas, {\'A}ngel and Huelga, Susana F and Plenio, Martin B},
  month      = sep,
  year       = {2014},
  pages      = {094001}
}

@article{josefsson_quantum-dot_2018,
  title     = {A quantum-dot heat engine operating close to the thermodynamic efficiency limits},
  volume    = {13},
  copyright = {2018 The Author(s)},
  issn      = {1748-3395},
  url       = {https://www.nature.com/articles/s41565-018-0200-5},
  doi       = {10.1038/s41565-018-0200-5},
  language  = {en},
  number    = {10},
  urldate   = {2024-11-15},
  journal   = {Nature Nanotechnology},
  author    = {Josefsson, Martin and Svilans, Artis and Burke, Adam M. and Hoffmann, Eric A. and Fahlvik, Sofia and Thelander, Claes and Leijnse, Martin and Linke, Heiner},
  month     = oct,
  year      = {2018},
  pages     = {920--924}
}

@article{martino_quantum_1996,
  title    = {Quantum groups and thermo field dynamics},
  volume   = {10},
  issn     = {0217-9792},
  url      = {https://www.worldscientific.com/doi/abs/10.1142/S0217979296000702},
  doi      = {10.1142/S0217979296000702},
  number   = {13n14},
  urldate  = {2024-11-24},
  journal  = {Int. J. Mod. Phys. B},
  author   = {Martino, S. De and Siena, S. De and Vitiello, G.},
  month    = jun,
  year     = {1996},
  pages    = {1615--1624}
}

@article{woods_mappings_2014,
  title    = {Mappings of open quantum systems onto chain representations and {Markovian} embeddings},
  volume   = {55},
  issn     = {0022-2488},
  url      = {https://doi.org/10.1063/1.4866769},
  doi      = {10.1063/1.4866769},
  number   = {3},
  urldate  = {2025-04-04},
  journal  = {J. Math. Phys.},
  author   = {Woods, M. P. and Groux, R. and Chin, A. W. and Huelga, S. F. and Plenio, M. B.},
  month    = mar,
  year     = {2014},
  pages    = {032101}
}

@article{celeghini_thermo_1998,
  title     = {Thermo field dynamics and quantum algebras},
  volume    = {244},
  copyright = {https://www.elsevier.com/tdm/userlicense/1.0/},
  issn      = {03759601},
  url       = {https://linkinghub.elsevier.com/retrieve/pii/S0375960198004472},
  doi       = {10.1016/S0375-9601(98)00447-2},
  language  = {en},
  number    = {6},
  urldate   = {2025-04-04},
  journal   = {Physics Letters A},
  author    = {Celeghini, E. and De Martino, S. and De Siena, S. and Iorio, A. and Rasetti, M. and Vitiello, G.},
  month     = aug,
  year      = {1998},
  pages     = {455--461}
}

@article{bulla_numerical_2003,
  title     = {Numerical {Renormalization} {Group} for {Bosonic} {Systems} and {Application} to the {Sub}-{Ohmic} {Spin}-{Boson} {Model}},
  volume    = {91},
  copyright = {http://link.aps.org/licenses/aps-default-license},
  issn      = {0031-9007, 1079-7114},
  url       = {https://link.aps.org/doi/10.1103/PhysRevLett.91.170601},
  doi       = {10.1103/PhysRevLett.91.170601},
  language  = {en},
  number    = {17},
  urldate   = {2025-04-04},
  journal   = {Phys. Rev. Lett.},
  author    = {Bulla, Ralf and Tong, Ning-Hua and Vojta, Matthias},
  month     = oct,
  year      = {2003},
  pages     = {170601}
}

@article{leggett_dynamics_1987,
  title     = {Dynamics of the dissipative two-state system},
  volume    = {59},
  copyright = {http://link.aps.org/licenses/aps-default-license},
  issn      = {0034-6861},
  url       = {https://link.aps.org/doi/10.1103/RevModPhys.59.1},
  doi       = {10.1103/RevModPhys.59.1},
  language  = {en},
  number    = {1},
  urldate   = {2025-04-02},
  journal   = {Rev. Mod. Phys.},
  author    = {Leggett, A. J. and Chakravarty, S. and Dorsey, A. T. and Fisher, Matthew P. A. and Garg, Anupam and Zwerger, W.},
  month     = jan,
  year      = {1987},
  pages     = {1--85}
}

@article{zwolak_mixed-state_2004,
  title      = {Mixed-{State} {Dynamics} in {One}-{Dimensional} {Quantum} {Lattice} {Systems}: {A} {Time}-{Dependent} {Superoperator} {Renormalization} {Algorithm}},
  volume     = {93},
  copyright  = {http://link.aps.org/licenses/aps-default-license},
  issn       = {0031-9007, 1079-7114},
  shorttitle = {Mixed-{State} {Dynamics} in {One}-{Dimensional} {Quantum} {Lattice} {Systems}},
  url        = {https://link.aps.org/doi/10.1103/PhysRevLett.93.207205},
  doi        = {10.1103/PhysRevLett.93.207205},
  language   = {en},
  number     = {20},
  urldate    = {2025-01-26},
  journal    = {Phys. Rev. Lett.},
  author     = {Zwolak, Michael and Vidal, Guifré},
  month      = nov,
  year       = {2004},
  pages      = {207205}
}

@article{verstraete_matrix_2004,
  title      = {Matrix {Product} {Density} {Operators}: {Simulation} of {Finite}-{Temperature} and {Dissipative} {Systems}},
  volume     = {93},
  copyright  = {http://link.aps.org/licenses/aps-default-license},
  issn       = {0031-9007, 1079-7114},
  shorttitle = {Matrix {Product} {Density} {Operators}},
  url        = {https://link.aps.org/doi/10.1103/PhysRevLett.93.207204},
  doi        = {10.1103/PhysRevLett.93.207204},
  language   = {en},
  number     = {20},
  urldate    = {2025-01-26},
  journal    = {Phys. Rev. Lett.},
  author     = {Verstraete, F. and García-Ripoll, J. J. and Cirac, J. I.},
  month      = nov,
  year       = {2004},
  pages      = {207204}
}

@article{levy_quasiprobability_2020,
  title    = {Quasiprobability {Distribution} for {Heat} {Fluctuations} in the {Quantum} {Regime}},
  volume   = {1},
  issn     = {2691-3399},
  url      = {https://link.aps.org/doi/10.1103/PRXQuantum.1.010309},
  doi      = {10.1103/PRXQuantum.1.010309},
  language = {en},
  number   = {1},
  urldate  = {2024-12-02},
  journal  = {PRX Quantum},
  author   = {Levy, Amikam and Lostaglio, Matteo},
  month    = sep,
  year     = {2020},
  pages    = {010309}
}

@article{segal2005spin,
  title     = {Spin-boson thermal rectifier},
  author    = {Segal, Dvira and Nitzan, Abraham},
  journal   = {Physical review letters},
  volume    = {94},
  number    = {3},
  pages     = {034301},
  year      = {2005},
  publisher = {APS},
  url       = {https://journals.aps.org/prl/abstract/10.1103/PhysRevLett.94.034301}
}

@article{saito2013kondo,
  title     = {Kondo signature in heat transfer via a local two-state system},
  author    = {Saito, Keiji and Kato, Takeo},
  journal   = {Physical review letters},
  volume    = {111},
  number    = {21},
  pages     = {214301},
  year      = {2013},
  publisher = {APS},
url={https://journals.aps.org/prl/abstract/10.1103/PhysRevLett.111.214301}
}

@article{wang2015multilayer,
  title     = {Multilayer multiconfiguration time-dependent Hartree theory},
  author    = {Wang, Haobin},
  journal   = {The Journal of Physical Chemistry A},
  volume    = {119},
  number    = {29},
  pages     = {7951--7965},
  year      = {2015},
  publisher = {ACS Publications},
url={https://pubs.acs.org/doi/10.1021/acs.jpca.5b03256#Abstract}
}

@article{nam_diagonalization_2016,
  title    = {Diagonalization of bosonic quadratic {Hamiltonians} by {Bogoliubov} transformations},
  volume   = {270},
  issn     = {0022-1236},
  url      = {https://www.sciencedirect.com/science/article/pii/S0022123615004905},
  doi      = {10.1016/j.jfa.2015.12.007},
  number   = {11},
  urldate  = {2024-11-21},
  journal  = {Journal of Functional Analysis},
  author   = {Nam, Phan Thành and Napiórkowski, Marcin and Solovej, Jan Philip},
  month    = jun,
  year     = {2016},
  pages    = {4340--4368}
}

@article{das_supersymmetry_1989,
  title    = {Supersymmetry and finite temperature},
  volume   = {158},
  issn     = {0378-4371},
  url      = {https://www.sciencedirect.com/science/article/pii/0378437189905025},
  doi      = {10.1016/0378-4371(89)90502-5},
  number   = {1},
  urldate  = {2025-01-27},
  journal  = {Physica A: Stat. Mech. App.},
  author   = {Das, Ashok},
  month    = may,
  year     = {1989},
  pages    = {1--21}
}

@article{bogoljubov_new_1958,
  title     = {A {New} {Method} in the {Theory} of {Superconductivity}},
  volume    = {6},
  copyright = {Copyright © 1958 WILEY-VCH Verlag GmbH \& Co. KGaA, Weinheim},
  issn      = {1521-3978},
  url       = {https://onlinelibrary.wiley.com/doi/abs/10.1002/prop.19580061102},
  doi       = {10.1002/prop.19580061102},
  language  = {de},
  number    = {11-12},
  urldate   = {2025-01-27},
  journal   = {Fortschritte der Physik},
  author    = {Bogoljubov, N. N. and Tolmachov, V. V. and Širkov, D. V.},
  year      = {1958},
  pages     = {605--682}
}

@misc{mcculloch_infinite_2008,
  author        = {McCulloch, I. P.},
  year          = {2008},
  month         = apr,
  number        = {arXiv:0804.2509},
  eprint        = {0804.2509},
  primaryclass  = {cond-mat},
  publisher     = {arXiv},
  urldate       = {2025-04-30},
  archiveprefix = {arXiv}
}

@article{balian_nonunitary_1969,
  title    = {Nonunitary bogoliubov transformations and extension of {Wick}’s theorem},
  volume   = {64},
  issn     = {1826-9877},
  url      = {https://doi.org/10.1007/BF02710281},
  doi      = {10.1007/BF02710281},
  language = {en},
  number   = {1},
  urldate  = {2025-01-03},
  journal  = {Il Nuovo Cimento B (1965-1970)},
  author   = {Balian, R. and Brezin, E.},
  month    = nov,
  year     = {1969},
  pages    = {37--55}
}

@article{takahashi_thermo_1996,
  title    = {Thermo field dynamics},
  volume   = {10},
  doi      = {10.1142/S0217979296000817},
  journal  = {Int. J. Mod. Phys. B},
  author   = {Takahashi, Y. and Umezawa, H.},
  year     = {1996},
  pages    = {1755--1805}
}

@article{strathearn_efficient_2018,
  title     = {Efficient non-{Markovian} quantum dynamics using time-evolving matrix product operators},
  volume    = {9},
  copyright = {2018 The Author(s)},
  issn      = {2041-1723},
  url       = {https://www.nature.com/articles/s41467-018-05617-3},
  doi       = {10.1038/s41467-018-05617-3},
  language  = {en},
  number    = {1},
  urldate   = {2025-01-03},
  journal   = {Nature Communications},
  author    = {Strathearn, A. and Kirton, P. and Kilda, D. and Keeling, J. and Lovett, B. W.},
  month     = aug,
  year      = {2018},
  pages     = {3322}
}

@article{jorgensen_exploiting_2019,
  title    = {Exploiting the {Causal} {Tensor} {Network} {Structure} of {Quantum} {Processes} to {Efficiently} {Simulate} {Non}-{Markovian} {Path} {Integrals}},
  volume   = {123},
  url      = {https://link.aps.org/doi/10.1103/PhysRevLett.123.240602},
  doi      = {10.1103/PhysRevLett.123.240602},
  number   = {24},
  urldate  = {2025-01-03},
  journal  = {Phys. Rev. Lett.},
  author   = {Jørgensen, Mathias R. and Pollock, Felix A.},
  month    = dec,
  year     = {2019},
  pages    = {240602}
}

@article{cygorek_simulation_2022,
  title     = {Simulation of open quantum systems by automated compression of arbitrary environments},
  volume    = {18},
  copyright = {2022 The Author(s), under exclusive licence to Springer Nature Limited},
  issn      = {1745-2481},
  url       = {https://www.nature.com/articles/s41567-022-01544-9},
  doi       = {10.1038/s41567-022-01544-9},
  language  = {en},
  number    = {6},
  urldate   = {2025-03-08},
  journal   = {Nature Physics},
  author    = {Cygorek, Moritz and Cosacchi, Michael and Vagov, Alexei and Axt, Vollrath Martin and Lovett, Brendon W. and Keeling, Jonathan and Gauger, Erik M.},
  month     = jun,
  year      = {2022},
  pages     = {662--668}
}

@article{fishman_itensor_2022,
  title    = {The {ITensor} {Software} {Library} for {Tensor} {Network} {Calculations}},
  issn     = {2949-804X},
  url      = {https://www.scipost.org/SciPostPhysCodeb.4},
  doi      = {10.21468/SciPostPhysCodeb.4},
  language = {en},
  urldate  = {2025-04-11},
  journal  = {SciPost Physics Codebases},
  author   = {Fishman, Matthew and White, Steven and Stoudenmire, Edwin Miles},
  month    = aug,
  year     = {2022},
  pages    = {004}
}
\appendix

\counterwithout{equation}{section}
\counterwithout{figure}{section}
\counterwithout{table}{section}

\setcounter{equation}{0}
\setcounter{figure}{0}
\setcounter{table}{0}
\setcounter{page}{7}

\renewcommand{\theequation}{S\arabic{equation}}
\renewcommand{\thefigure}{S\arabic{figure}}
\renewcommand{\thetable}{S\arabic{table}}

\renewcommand{\bibnumfmt}[1]{[S#1]}
\renewcommand{\citenumfont}[1]{S#1}

\widetext
\clearpage
\begin{center}
\textbf{\large Supplemental Materials for ``Heat operator approach to quantum stochastic thermodynamics in the strong-coupling regime"}
\end{center}

\section{Derivation of Eq.\ (\ref{eq:chiu_q})}
\label{app:chiu_q}
We assume an initial product state of the system and the bath, $\hat{\rho}(0) = \hat\rho _{\text{\tiny $S$}}(0) \otimes \hat \pi_{\beta} $ where the thermal state $\hat\pi_{\beta}$ is of the form \eqref{eq:trace_tau}. In the 2-point measurement scheme, the probability distribution of fluctuations of heat ($Q(t)$) is given by
\begin{equation*}
     P(Q,t)  = \sum_{m,n} \text{Tr} \{ \Pi_{n}\mathcal{U}_t[\Pi_{m} \hat{\rho}(0)\Pi_{m}]\}\delta (Q + E_{m} - E_{n}).
\end{equation*}
Thus the characteristic function of heat is given by 
\begin{align}
\chi(\lambda,t) &= \int dQ P(Q)\, e^{i\lambda Q} \\
&= \sum_{m,n} \text{Tr} \{ \Pi_{n}\mathcal{U}_t[\Pi_{m} \hat{\rho}(0)\Pi_{m}]\}\, e^{i\lambda(E_n-E_m)} \\
&=  \text{Tr} \{\sum_{n}e^{i\lambda E_n} \Pi_{n}\, \mathcal{U}_t[\sum_{m} e^{-i \lambda E_m} \Pi_{m}\, \hat\rho _{\text{\tiny $S$}}(0) \otimes \hat\pi_{\beta}\, \Pi_{m}]\} \\
&=  \text{Tr} \{\sum_{n}e^{i\lambda E_n} \Pi_{n}\, \mathcal{U}_t[\sum_{m} e^{-i\lambda E_m} \Pi_{m} \hat{\rho}(0)]\} \\
&= \text{Tr}\left\{e^{i\lambda\hat H_{B}} \,\mathcal{U}_t[e^{-i\lambda\hat H_{B}} \hat{\rho}(0) ]\right\}
\end{align}

\section{Verification of Eq.\ (\ref{eq:main_result})}
\label{app:state_proof}
The key ingredient required in characterizing fluctuating heat exchange is the characteristic function denoted by $\chi (\lambda,t)$, that has the standard form in  \eqref{eq:chiu_q}. In this section, we begin with our proposed modified characteristic function expression \eqref{eq:simpl_chi_q} and derive \eqref{eq:chiu_q}. This, in turn, will prove the primary statement of the article stated in Eq.\ \eqref{eq:main_result}.  \\

We start by expanding \eqref{eq:simpl_chi_q}: 
\begin{eqnarray}
    \chi (\lambda,t)  &=& \text{Tr}\left\{e^{i\lambda\Tilde{\mathcal{Q}}}\,\mathcal{U}^{\text{\tiny $G$}}_t[\hat\rho _{\text{\tiny $S$}}(0) \otimes \Tilde{\Phi}]\right\} \nonumber \\
    &=& \text{Tr}\left\{e^{i\lambda\Tilde{\mathcal{Q}}} \, \tilde{U}_{\text{\tiny $G$}}(t)\, (\hat\rho _{\text{\tiny $S$}}(0) \otimes \Tilde{\Phi})  \, \tilde{U}_{\text{\tiny $G$}}^\dagger(t) \,\right\}  \qquad  \nonumber \\
    &=& \text{Tr}\left\{e^{i\lambda \Tilde{\mathcal{Q}}}\, \tilde{U}_{\text{\tiny $G$}}(t) \, e^{-i\lambda \Tilde{\mathcal{Q}}}\,  (\hat\rho _{\text{\tiny $S$}}(0) \otimes \Tilde{\Phi})  \, \tilde{U}^\dagger_{\text{\tiny $G$}}(t) \,\right\} \qquad \left[\because \Tilde{\mathcal{Q}}(\hat\rho _{\text{\tiny $S$}}(0) \otimes \Tilde{\Phi})=0\right] \nonumber\\
    &=& \text{Tr}\left\{e^{i\lambda \Tilde{\mathcal{Q}}}\, \mathcal{G}_{\text{\tiny$\beta$}}[ \tilde{U}(t)] \, e^{-i\lambda \Tilde{\mathcal{Q}}}\, (\hat\rho _{\text{\tiny $S$}}(0) \otimes \Tilde{\Phi} )\, \mathcal{G}_{\text{\tiny$\beta$}}[\tilde{U}^\dagger(t)]  \,\right\},
\end{eqnarray}
where $\tilde{U}(t) = \overset{\leftarrow}{\rm T}{\rm exp}[-i\int_0^t dt'\, \tilde{H}(t')]$ is the time-evolution operator generated by $\tilde{H}(t)$, as defined in Eq.~\eqref{eq:HTF}. Note that $\hat G_\text{\tiny$\beta$}= - i \sum \nolimits _{\nu} \theta_{\text{\tiny$\beta$},\nu} (\hat{a}_{\nu}^{\dagger}\hat{b}_{\nu}^{\dagger}-\hat{b}_{\nu}\hat{a}_{\nu})$ commutes with $\Tilde{\mathcal{Q}}$, i.e., $[\hat{G}_{\text{\tiny$\beta$}},\Tilde{\mathcal{Q}}]=0$. Then one can write for any operator $\hat x$, 
\begin{equation}
    e^{i\lambda\mathcal{\Tilde{Q}}}\mathcal{G}_{\text{\tiny$\beta$}}[\hat x]e^{-i\lambda\mathcal{\Tilde{Q}}}=\mathcal{G}_{\text{\tiny$\beta$}}[e^{i\lambda\mathcal{\Tilde{Q}}}\,\hat x\, e^{-i\lambda\mathcal{\Tilde{Q}}}].\label{eq:commutes}
\end{equation}

Using \eqref{eq:commutes} and the thermal vacuum state $\Tilde{\tau}_{\text{\tiny$\beta$}}= \mathcal{G}_{\text{\tiny$\beta$}}^{-1}[\Tilde{\Phi} ]$ from Eq.\ \eqref{eq:vacuum_OA}, we get  
\begin{eqnarray}    
    \chi (\lambda,t) &=& \text{Tr}\left\{\mathcal{G}_{\text{\tiny$\beta$}}[e^{i\lambda \Tilde{\mathcal{Q}}}\,  \tilde{U}(t) \, e^{-i\lambda \Tilde{\mathcal{Q}}}]\, (\hat\rho _{\text{\tiny $S$}}(0) \otimes \Tilde{\Phi} )\, \mathcal{G}_{\text{\tiny$\beta$}}[ \tilde{U}^\dagger(t)]  \,\right\} \label{l1}\\
    &=& \text{Tr}\left\{e^{i\lambda\Tilde{\mathcal{Q}}}\,\tilde{U}(t)\,e^{-i\lambda\Tilde{\mathcal{Q}}}\,(\hat\rho _{\text{\tiny $S$}}(0) \otimes \Tilde{\tau}_{\text{\tiny$\beta$}})\,\tilde{U}^\dagger(t)\right\} \label{l2}
\end{eqnarray}

In the following, we use $\Tilde{H}(t) =  \hat{H}(t) - \hat H_{B,A}$ (Eq.\ \eqref{eq:HTF}) and that $[\hat H, \hat H_{B,A}]=[\hat H_{B,O}, \hat H_{B,A}]=0$, from which it follows that $\tilde{U}(t) = \hat{U}(t) e^{i t\hat{H}_{B,A}} = e^{i t\hat{H}_{B,A}}\hat{U}(t)$, where $\hat{U}(t)=\overset{\leftarrow}{\rm T}{\rm exp}[-i\int_0^t dt'\, \hat{H}(t')]$ is generated by the original Hamiltonian $\hat{H}(t)$ in Eq.~\eqref{eq:original_H}. Finally, using Eq.~\eqref{eq:trace_tau} we trace out the auxiliary bath $A$ (\textcolor{newtext}{$\text{Tr}_O$ denotes trace operation over the system and the original bath $O$}), we recover the standard expression for heat characteristic function in Eq.\ \eqref{eq:chiu_q}.
    \begin{eqnarray}
     \chi (\lambda,t) &=& \text{Tr}\left\{ e^{i\lambda (\hat H_{B,O}-\hat H_{B,A})} \, \tilde{U}(t)   \, e^{-i\lambda (\hat H_{B,O}-\hat H_{B,A})} \, (\hat\rho _{\text{\tiny $S$}}(0) \otimes \Tilde{\tau}_{\text{\tiny$\beta$}} )\, \tilde{U}^\dagger(t)   \,\right\} \nonumber\\
     &=& \text{Tr}_O\left\{e^{i\lambda \hat H_{B,O}} \, \hat{U}(t) \, e^{-i\lambda \hat H_{B,O}}\,(\hat\rho _{\text{\tiny $S$}}(0) \otimes \text{Tr}_A\left\{\Tilde{\tau}_{\text{\tiny$\beta$}}\right\} )\, \hat{U}^\dagger(t) \,\right\} \nonumber\\
    &=& \text{Tr}_O\left\{e^{i\lambda \hat H_{B,O}} \,\hat{U}(t) \, e^{-i\lambda \hat H_{B,O}}\, (\hat\rho _{\text{\tiny $S$}}(0) \otimes \hat \pi_{\text{\tiny$\beta$}} ) \,  \hat{U}^\dagger(t) \, \right\} \nonumber\\
    &=& \text{Tr}_O\left\{e^{i\lambda\hat H_{B,O}} \,\mathcal{U}_t[e^{-i\lambda\hat H_{B,O}} \hat{\rho}(0) ]\right\},
\end{eqnarray}
which proves Eq.~\eqref{eq:simpl_chi_q}. Using the definition in Eq.\ \eqref{eq:mu_moments}, we can state that indeed the $n^{\rm th}$ moment of heat is given by,
\begin{equation}
       \langle Q ^n(t)\rangle = \text{Tr}\left\{\Tilde{\mathcal{Q}}^n \, \mathcal{U}^{\text{\tiny $G$}}_t[\hat\rho _{\text{\tiny $S$}}(0) \otimes \Tilde{\Phi}]\right\}. \label{eq:simpl_heat_char}
\end{equation}
This concludes the proof of our main result in Eq.\ \eqref{eq:main_result}.
\section{Time evolving density matrix using orthogonal polynomials algorithm (TEDOPA)}
\label{app:TEDOPA}
The crux of this method is an exact unitary transformation of any Gaussian bosonic or fermionic bath into a 1D chain of modes with nearest-neighbor coupling \hyperlink{suppref:tedopa}{[S1]}. This transformation is generated by the family of orthogonal polynomials determined by the couplings $g_{j\nu}$ entering the linear interaction Hamiltonian. While there exist other ways to discretize a continuum of modes, it was shown in \hyperlink{suppref:deVega}{[S2]}  that the orthogonal polynomial strategy is optimal for a quadratic bath Hamiltonian. A key feature of this method is that the environment's state is directly accessible.

Given that the environment comprises reservoirs, each characterized by a spectral density $J_{j}(\omega)$, TEDOPA presents an exact unitary transformation $U_{j,n}(\omega)$ mapping the continuum of bosonic or fermionic modes to discrete semi-infinite chains (each corresponding to a single bath) with only nearest-neighbor interactions. For each bath $B_j$, $U_{j,n}(\omega)=g_{j}(\omega) p_{j,n}(\omega)$ is defined such that $\int d\omega\, U_{j,m}(\omega)U_{j,n}(\omega) = \delta_{mn}$. That is, $\{p_{j,n}(\omega)\}_n$ is the set of orthogonal polynomials with respect to the measure $d\mu(\omega) = g_{j}^{2}(\omega) d\omega$. The system-bath coupling strengths $g_{j}(\omega)$ (see Eq.\eqref{eq:interaction_ham}) thus determines the chain-mapped bath operators

\begin{equation}
    \hat{b}_{j,n} =  \int d\omega\, U_{j,n}(\omega) \hat{a}_{j,\omega}.  \label{eq:chain-mapping}
\end{equation}
The chain-mapped bath operators $\{\hat{b}_{j,n}\}$ satisfy the standard commutation relations for the respective baths. The Hamiltonian (Eq.~\eqref{eq:original_H}) after chain mapping Eq.~\eqref{eq:chain-mapping} becomes, 
\begin{equation}
    \hat{H}_{CM} = \hat{H}_{S} + c_{0} \hat{A} (b_{0} +  b_{0}^{\dagger}) + \sum_{n = 0}^{\infty} \omega_{n}b_{n}^{\dagger}b_{n} + t_{n}(b_{n + 1}^{\dagger}b_{n} + b_{n}^{\dagger}b_{n + 1}).    \label{eq:tedopaham} 
\end{equation}
Here $c_{0}^2 = \int d\mu(\omega)$ is the coupling of the system with the first site of the chain-mapped bath. The tunneling frequencies $t_{n}$ and mode frequencies $\omega_{n}$ are the recurrence coefficients corresponding to the measure $d\mu(\omega)$.

\textcolor{newtext}{In this approach, the environments $O$ and $A$ are mapped onto independent 1D chains, where each chain mapping is generated by a family of orthogonal polynomials determined by the couplings ($u_{\beta,\nu} g_\nu$ or $v_{\beta,\nu} g_\nu$) entering the linear interaction Hamiltonian in Eq.~\eqref{transformed_Hamiltonian}. }


\section{Benchmark -- Independent spin-boson model}
\label{app:indep-model}
The total Hamiltonian of the independent spin-boson model is as follows:
\begin{equation}
    \hat{H} = \Delta \hat{S}_x  + \sum_{\nu} \omega_{\nu} \hat{a}^{\dagger}_{\nu} \hat{a}_{\nu} + \hat{S}_x \otimes \sum\nolimits_{\nu} g_{\nu} ( \hat{a}_{\nu} + \hat{a}^{\dagger}_{
\nu
} )
\end{equation}
We notice that the spin energy is conserved in this model, since $[\hat S_x, \hat H] = 0$. Thus, the initial state of the spin does not affect the heat transferred to the bath during the ensuing equilibration process. Furthermore, this model is exactly solvable, so that we can compute the heat characteristic function ($\chi(\lambda,t)$ in Eq.\ \eqref{eq:chiu_q}) and the heat moments analytically. It can be shown that \hyperlink{suppref:popovic}{[S3]}
\begin{equation}
    \ln \chi(\lambda, t) = -\frac{1}{2} \int _{0} ^{\infty} d\omega\, \frac{J(\omega)}{\omega^2} (1-\text{cos}(\omega t)) \Big( (1-\cos(\omega \lambda))\,\text{coth}\Big(\frac{\beta \omega}{2}\Big) - i \sin(\omega \lambda)\Big).
\end{equation}
From the characteristic function expression above, we can get the cumulants of heat distribution as
\begin{equation}
    \langle\langle Q^n(t) \rangle \rangle = (-i)^n \partial_\lambda ^{n} \ln \chi(\lambda, t)\big|_{\lambda = 0}.
\end{equation}

Specifically, the first and second moments are as follows:
\begin{eqnarray}
   &\ & \langle Q(t) \rangle = \frac{1}{2} \int _{0} ^{\infty} d\omega\, \frac{J(\omega)}{\omega} (1-\text{cos}(\omega t)) = \frac{\alpha \omega_C ^3 t^2}{1+\omega_C^2 t^2}
    \implies  \langle Q \rangle_{t=\infty} = \alpha \omega_C; \label{eq:mean_app}\\
    &\ & \langle\langle Q^2(t) \rangle\rangle = \langle Q^2(t) \rangle - \langle Q(t) \rangle^2 = \frac{1}{2} \int _{0} ^{\infty} d\omega\, J(\omega) (1-\text{cos}(\omega t)) \text{coth}\Big(\frac{\beta \omega}{2}\Big). \label{eq:var_app}
\end{eqnarray}
The expression for variance does not have a closed functional form, but it can be evaluated numerically.

\begin{figure}
    \centering
    \includegraphics[width=0.7\linewidth]{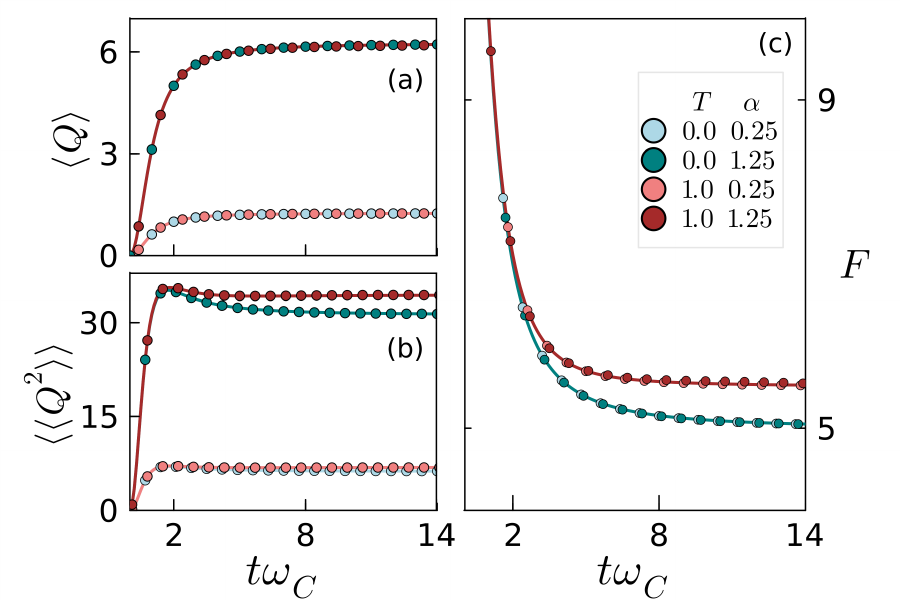}
    \caption{\textcolor{newtext}{Evolution of (a)~mean, (b)~variance, and (c)~Fano factor of heat fluctuations in the independent-boson model ($\epsilon_0=0,\ \Delta=1, \ \omega_c = 5$). The initial state of the spin is the $|+\rangle$ eigenstate of $\hat{S}_x$. Exact solutions for this model are plotted as solid lines. }}    
    \label{fig:exact_model}
\end{figure}

Fig.\ \ref{fig:exact_model} shows the time evolution of the mean $\langle Q(t)\rangle$ and variance $\langle \langle Q^2(t)\rangle \rangle$ for the independent-boson model, taking the initial state $\hat{\rho}_S(0) = |+\rangle\langle +|$, where $\hat{S}_x|+\rangle = |+\rangle$. The heat statistics in this model are independent of the initial spin state~\hyperlink{suppref:popovic}{[S3]}. We plot the behavior of this exactly solvable model as shown in Eqs.\ \eqref{eq:mean_app} and \eqref{eq:var_app}, to compare with the numerical results obtained using our proposed scheme for computing heat fluctuations. This model thus serves to benchmark our method and we see excellent agreement with the exact solutions.


\section{Computational complexity}
\label{app:complexity}

\textcolor{newtext}{Calculating heat fluctuations using the heat operator approach Eq.~\eqref{eq:main_result} presented in this article requires a standard MPS time evolution followed by the calculation of expectation value of the $n$-th power of the heat operator $\Tilde{\mathcal{Q}}$. The numerical implementation combines the heat-operator formalism with a chain mapping of each Gaussian bath into a 1D tensor network state. The primary computational cost is therefore determined by the real-time evolution of a one-dimensional chain of system and bath sites (see Eq.~\eqref{eq:tedopaham}) and calculation of the expectation value of the powers of the heat operator $\Tilde{\mathcal{Q}}$. In principle, for relatively small Hilbert spaces, we can use exact diagonalization or Krylov subspace methods [\hyperlink{suppref:kosloff1988time}{S4}, \hyperlink{suppref:hochbruck1997krylov}{S5}, \hyperlink{suppref:wang2003multilayer}{S6}] for the unitary time evolution of the initial system-bath vacuum state although here we will mainly consider macroscopic baths. }

\textcolor{newtext}{For a system $S$ consisting of $L_S$ sites and with $S$ being coupled to $N_{B}$
number of baths, all of which are mapped to a chain of length $L_{j}$ with local dimension $d_{B,j}$, the total chain length is
\begin{equation}
    L_{tot} = L_{S} + 2 \sum_{j = 1}^{N_{B}} L_{j},
\end{equation}
where the factor of 2 reflects the thermofield doubling of each bath. The time evolution can be carried out using a nearest–neighbour Trotter decomposition and MPS algorithms. e.g., time evolving block decimation (TEBD) or time-dependent variational principle (TDVP). Given a bond dimension $D$ and physical dimension per site $d$, a two–site update in TEBD or TDVP involves a singular value decomposition (SVD) of an effective matrix of dimensions $d\, D \times d\,D$, leading to a local cost of the order of $\mathcal{O}(d^3\, D^3)$~\hyperlink{suppref:paeckel}{[S7]}. Therefore, the overall cost per time step scales as $\mathcal{O}(L_{tot} \,d_{\text{max}}^3\, D^3)$,  with $d_{\text{max}} = \text{max}\{d_{S}, d_{B,j}\}$, up to constant factors that depend on the precise truncation strategy. The memory cost of storing the MPS scales as $\mathcal{O} (L_{\mathrm{tot}} d_{\max}\, D^2)$, while the total cost to reach a final time $t_{\max} = N_t \Delta t$ is of the order of $ \mathcal{O} ( N_t\, L_{\mathrm{tot}} d_{\max}^3\, D^3)$. In TEDOPA, the bath memory time is encoded in the chain lengths $L_j$ (chosen such that excitations do not reach the boundaries within $t_{\max}$), rather than in a tensor of size set by an explicit memory depth $K$ as in influence-functional approaches, so the cost grows linearly with the effective memory time at fixed $D$. In generic non-integrable systems, the required bond dimension increases with time due to entanglement growth; in the regimes studied here this growth remains moderate, and the total computational cost is kept under control by standard convergence checks ($D$, $L_j$, and $\Delta t$).}

\textcolor{newtext}{The heat operator $\tilde{Q}$ is a sum of bath Hamiltonians on the physical and auxiliary chains, and has bond dimension $D_{\tilde{Q}}$ that is independent of $L_{\mathrm{tot}}$. For a system with local Hilbert-space dimension $d_{\max}$ and MPS bond dimension $D$, 
a single sweep that contracts the tensor network $\langle \tilde{Q}^n(t)\rangle$
by updating left (or right) environments at each site has a computational cost that scales as 
$\mathcal{O}\!\left(L_{\mathrm{tot}} d_{\max}^2 D^2 D_{\tilde{Q}^n}\right)$ and with $D_{\tilde{\mathcal{Q}}} = \mathcal{O}(1)$ for quadratic bath Hamiltonians, it constitutes only a polynomial overhead on top of the complexity of the order of $ \mathcal{O} ( N_t\, L_{\mathrm{tot}} d_{\max}^3\, D^3)$ for the
underlying MPS time evolution for a given time $t$, as shown above.  }

\par\vskip18pt
\centerline{%
  \rule{0.12\textwidth}{0.5pt}\hskip -1.25em%
  \rule{0.12\textwidth}{1pt}\hskip -1.25em%
  \rule{0.12\textwidth}{1.25pt}\hskip -1.25em%
  \rule{0.12\textwidth}{1pt}\hskip -1.25em%
  \rule{0.12\textwidth}{0.5pt}%
}
\par\vskip18pt
\noindent
{\small \noindent\hypertarget{suppref:tedopa}{[S1]}  J. Prior, A. W. Chin, S. F. Huelga, and M. B. Plenio,
\href{https://doi.org/10.1103/PhysRevLett.105.050404}{Phys. Rev. Lett. 105, 050404 (2010)}.}\\
\small \noindent\hypertarget{suppref:deVega}{[S2]}  I. de Vega, U. Schollwöck, and A. F. Wolf, \href{https://link.aps.org/doi/10.1103/PhysRevB.92.155126}{Phys. Rev. B 92, 155126 (2015)}.\\
\small \noindent\hypertarget{suppref:popovic}{[S3]} M. Popovic, M. T. Mitchison, A. Strathearn, B. W. Lovett, J. Goold, and P. R. Eastham, \href{https://link.aps.org/doi/10.1103/PRXQuantum.2.020338}{PRX Quantum 2, 020338 (2021)}.\\
\small \noindent\hypertarget{suppref:kosloff1988time}{[S4]} R. Kosloff, \href{https://pubs.acs.org/doi/abs/10.1021/j100319a003}{The Journal of Physical Chemistry 92 (1988)}.\\
\small \noindent\hypertarget{suppref:hochbruck1997krylov}{[S5]} M. Hochbruck and C. Lubich, \href{https://epubs.siam.org/doi/10.1137/S0036142995280572}{SIAM Journal on Numerical Analysis 34.5 (1997)}.\\
\small \noindent\hypertarget{suppref:wang2003multilayer}{[S6]} H. Wang and M. Thoss, \href{https://pubs.aip.org/aip/jcp/article-abstract/119/3/1289/186240/Multilayer-formulation-of-the-multiconfiguration?redirectedFrom=fulltext}{J. Chem. Phys. 119 (2003)}.\\
\small \noindent\hypertarget{suppref:paeckel_time-evolution_2019}{[S7]} S. Paeckel, T. Köhler, A. Swoboda, S. R. Manmana, U. Schollwöck, and C. Hubig, \href{https://www.sciencedirect.com/science/article/pii/S0003491619302532}{Annals of Physics 411 (2019)}.

\end{document}